\documentclass[prb,twocolumn,showpacs,amsmath,amssymb,superscriptaddress]{revtex4-1}

\usepackage{amsmath,amssymb,amsfonts,bm,color,graphicx,tabularx,physics}

\usepackage[unicode=true,colorlinks=true]{hyperref}

\usepackage[etex=true,export]{adjustbox}

\hypersetup{linkcolor=blue,citecolor=blue,urlcolor=blue}
\usepackage{tabularx,multirow,array,diagbox}
\usepackage{adjustbox}

\newcommand{\beq}{\begin{equation}}
\newcommand{\eeq}{\end{equation}}
\newcommand{\beql}{\begin{equation*}}
\newcommand{\eeql}{\end{equation*}}
\newcommand{\beqn}{\begin{eqnarray}}
\newcommand{\eeqn}{\end{eqnarray}}
\newcommand{\nn}{\nonumber\\}

\begin{document}

\title{Intrinsic Andreev $\pi$-reflection and Josephson $\pi$-junction for centrosymmetric spin-triplet superconductors}
\author{Han-Bing Leng}
\author{Chuang Li} 
\author{Xin Liu}
\affiliation{School of Physics and Wuhan National High Magnetic Field Center, Huazhong University of
Science and Technology, Wuhan, Hubei 430074, China}
\date{\today}

\begin{abstract}
In this work, we systematically study two phases, called Andreev $\pi$-phase and orbital-phase, and their influence on the Josephson effect. When the system is time-reversal invariant and centrosymmetric, these two phases only appear in the odd-parity pairings. The Andreev $\pi$-phase has nothing to do with the specific form of the odd-parity pairings and means an intrinsic $\pi$-phase between the spin-triplet Cooper pairs entering and leaving CTSCs in the Andreev reflections. The orbital-phase corresponds to the phase difference between the spin-triplet Cooper pairs with opposite spin polarization and depends on the specific form of the odd-parity gap functions. When the normal region of the Josephson junction contacts the same side of the CTSCs with some specific odd-parity parings, the competition between the two phases can lead to the Josephson $\pi$-junction. Note that this junction is different from that of the conventional Josephson junction (JJ) and is dubbed a U-shaped junction according to its geometry. Meanwhile, in a conventional JJ, the interplay of these two phases causes their impact on the CPR to be completely canceled out. Therefore no matter what kind of pairing symmetries the CTSC has, it will lead to Josephson 0-junction in this case. We obtain our results based on the model of the M$_{x}$Bi$_2$Se$_3$ family where M may be Cu, Sr, or Nb. Therefore, we propose to detect the pairing symmetry of M$_{x}$Bi$_2$Se$_3$ through a superconducting quantum interference device containing a U-shaped Josephson junction.

\end{abstract}

\maketitle

\section{Introduction}
Current-phase relation is one of the most basic properties of the Josephson effect. In earlier studies \cite{Josephson1962,O.Kulik1969}, it took the form of $I_{\rm s}=I_{\rm c}\sin\phi$ with $I_{\rm c}$ the critical current and $\phi$ the phase difference caused by the flux enclosed in the superconducting circuit. The free energy of this conventional Josephson junction has a minimum at $\phi=0$, which is dubbed a $0$-junction. Later studies found that the CPR of Josephson effect is not limited to this sinusoidal form but also has very rich forms due to the various superconducting pairing symmetries \cite{Kontos2002,Tanaka2007,Lutchyn2010,Burmistrova2015}. Especially, the CPR and the associated free energy minimum can have a $\pi$ phase shift, resulting in the so-called $\pi$-junction \cite{Geshkenbein1987,Liu1995,Barash2002,Liu2016}. In the Josephson junction involving only spin-singlet pairings, the $\pi$-phase shift in the ground state can be simply understood as the sign change of the superconducting gap owing to the orbital symmetry of the Cooper pairs. As a result, the observations of the $\pi$-junction in the phase-sensitive measurements of the cuprate superconductor(SC) provide the smoking gun of the $d$-wave superconductivity \cite{Tsuei1994,VanHarlingen1995,Ilichev2001,Inomata2005,Kirshenbaum2013,Lu2015}. While, in the Josephson junction involving spin-triplet pairings, not only the phase but also the various spin structures \cite{Asano2003,Khaire2010,Brydon2011} of the order parameters can affect the CPR, which complicates the mechanism of the Josephson $\pi$-junction. For example, the $\pi$-junction has been observed in the conventional superconducting/ferromagnetic hybrids with the even-parity odd-frequency and spin-triplet pairings, which are caused by Zeeman splitting in the junction region and have nothing to do with the odd-parity superconductivity \cite{Bergeret2001,Volkov2003,Bergeret2003,Buzdin2003a,Asano2007a,Asano2007b,Brydon2009,liu2014a}. This indicates that the verification of the odd-parity superconductivity through phase-sensitive measurements cannot simply replicate the successful experimental scheme in $d$-wave superconductors. 

On the other hand, with the rapid development of topological superconductivity studies in the past decade, there have been some promising candidate materials for odd-parity superconductivity. For example, recent observations of spin-rotational symmetry breaking in the  centrosymmetric and time-reversal-invariant M$_x$Bi$_2,$Se$_3$, with M=Cu, Nb or Sr, seem to indicate the interorbital, odd-parity pairing \cite{Hashimoto2013,Matano2016,Pan2016,Trang2016,Asaba2017,Du2017,Schmidt2020}. This so-called nematic superconductivity suggests that the Cooper pairs consist of two electrons from different orbits in the Bi$_2$Se$_3$ low-energy structure \cite{Fu2010,Fu2014}, in contrast to standard intraorbital pairings. However, all of the current experiments measured the magnitude of gap function which alone cannot provide a convincing determination of the pairing symmetry. Thus it is necessary to determine the parity of superconducting pairings through appropriate phase-sensitive experiments.

\begin{figure}
\includegraphics[width=1.0\columnwidth]{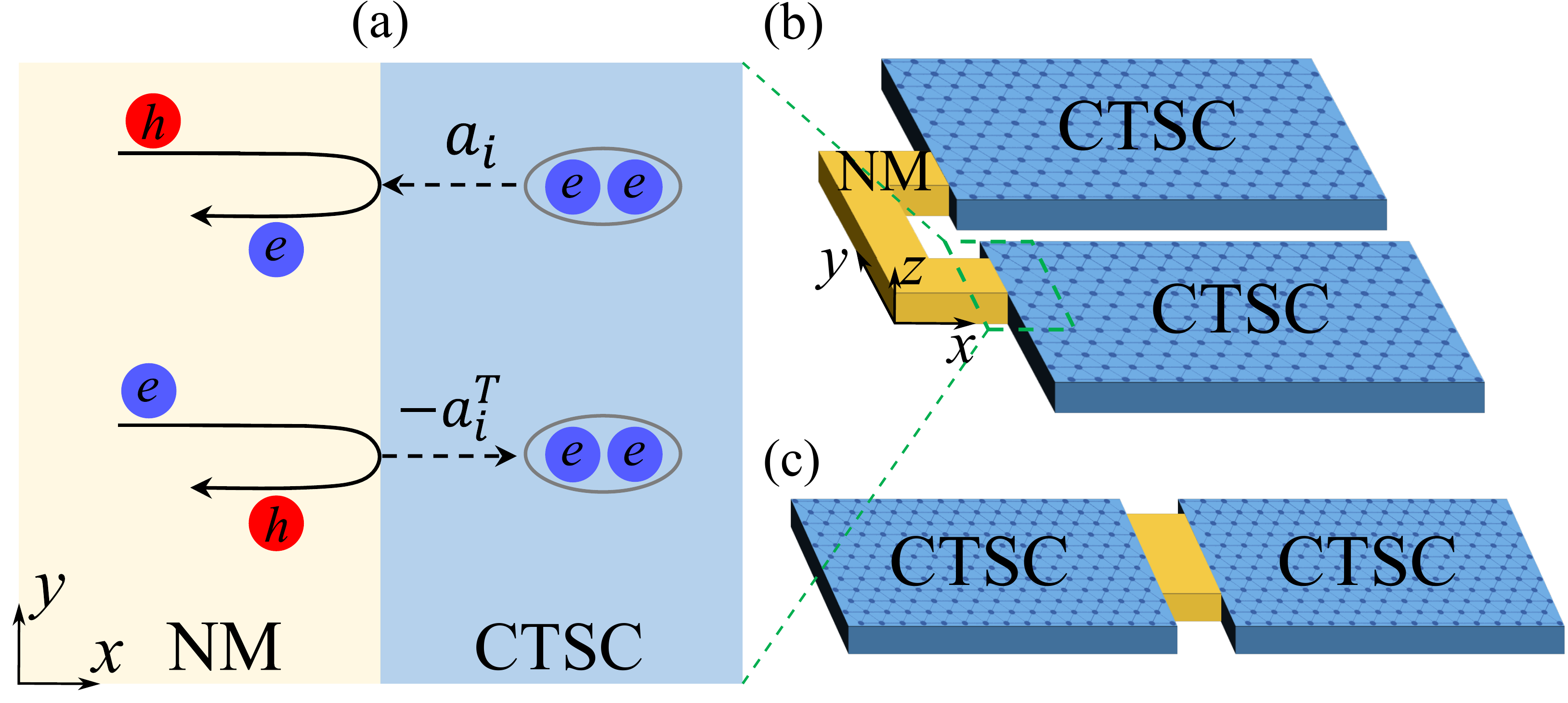}
\caption{(a) Andreev reflection at NM/CTSC interface. The Cooper pairs tunneling out of and into CTSC with Andreev reflection matrix $a_i$ and $-a_{i}^{\rm T}$ respectively. (b) A sketch of the U-shaped CTSC/NM/CTSC junction, in which its normal region contacts the same sides of the CTSCs. (c) A sketch of conventional Josephson junction.}
\label{junctions}
\end{figure}

In this work, we study the Josephson effect in CTSC with odd-parity superconducting order parameters. Microscopically the Josephson current is carried by the Andreev bound states (ABSs)\cite{O.Kulik1969}, which is formed by the quantum interference of multiple Andreev reflections in the junction. Therefore we start with a general symmetry analysis of the scattering matrix at the normal-metal(NM)/CTSC interface. We find that there is a time-reversal symmetry protected $\pi$-phase difference between the spin-triplet Cooper pairs entering and leaving the CTSC (Fig.~\ref{junctions}(a)). This Andreev $\pi$-phase is independent of the specific spin-triple forms. Meanwhile, there is an orbital-phase in the spin-triplet Andreev reflection channel, which stems from the relative phase difference between the spin-triplet pairs with the opposite spins. It can be seen from the scattering matrix that the orbital phase leads to perfect Andreev reflections under special energy, which corresponds exactly to the surface state energy, implying the physical reality of the orbital phase. Interestingly, when the normal region of the JJ contacts the same side of the superconductors (Fig.~\ref{junctions}(b)), referred to as the U-shaped junction according to its geometry, the CPR of this JJ shows the competition between these two phases. In the normal incidence channel, the CPR corresponds to the Josephson $\pi$-junction. In the process of changing from normal incidence to parallel incidence, the orbital-phase gradually changes the CPR from $\pi$-junction to $0$-junction. When the junction takes the traditional geometry structure, its normal region contacting the opposite sides of the superconductor (Fig.~\ref{junctions}(c)), the competition of these two phases in CPR completely disappears, which gives rise to the Josephson $0$-junction in all channels \cite{Tanaka1999,Asano2006,Asano2007c,Manikandan2017}. As a result, the CTST/NM/CTST junction in Fig.~\ref{junctions}(c) will perform as a conventional $0$-junction no matter what the superconducting order parameter is. Therefore we propose the phase-sensitive measurement in Josephson junction, with its NM contacting the same sides of the superconductor, to detect the odd-parity superconductivity.

This work is organized as follows. In Sec II, we study the Andreev reflection at the NM/CTSC interface. We first present a model-independent discussion of Andreev reflection based on symmetry analysis and show that there is a universal $\pi$ phase difference between the spin-triplet Cooper pair entering and leaving the CTSC. We then study the Andreev reflection in M$_x$Bi$_2$Se$_3$ for three possible odd-parity gap functions and confirm the existence of the Andreev-$\pi$ phase and orbital-phase both analytically and numerically. In Sec III, we first provide an intuitive picture of the interplay between the two phases in the Josephson effect. Especially we schematically show how the competition between the two phases exists and disappears in the junction of Fig.~\ref{orbital phase}(c) and (d) respectively. We then calculate both analytically and numerically the Andreev levels and CPRs in the U-shaped junction. In Sec IV, we apply our analysis in the Josephson effect of M$_x$Bi$_2$Se$_3$/NM/M$_x$Bi$_2$Se$_3$ junction and give the numerical results. In Sec V, we discuss the possible experimental realization of our proposals.

\section{Two special phases in spin-triplet Andreev reflection channels at NM/CTSC interface}\label{AR}

\subsection{Intrinsic $\pi$-phase in Andreev reflection matrix}\label{pi}
To consider the influence of various phases on the Josephson effect, it is convenient to start from studying the Andreev reflections\cite{Andreev1964} of the quasi-particles confined in the junction region. When injecting a quasi-particle from the normal lead to the superconductor, the reflection matrix of the NM/SC interface takes the form
\beqn\label{r-matrix}
R(E)=\left(\begin{array}{cc} r_{\rm ee} & r_{\rm eh} \\ r_{\rm he} & r_{\rm hh} \end{array}\right),
\eeqn
where the diagonal block $r_{\rm ee(hh)}$ describes the normal reflection of electron(hole)-like quasi-particles and the off-diagonal block $r_{\rm eh(he)}$ describing the Andreev reflection. Note that the reflection matrix describes the coupling between the transverse modes injected to and reflected by the NM/SC interface from the lead. If the lead has both time-reversal and inversion symmetry, the transverse modes at the Fermi level are doubly degenerate so that each block in Eq.~\eqref{r-matrix} is a $2N\times2N$ matrix with $2N$ the number of the transverse modes in the normal side and the prefactor 2 stems from spin. In this case, the reflection matrix can be further written as
\beqn\label{R-matrix}
R(E)=\left(\begin{array}{cc} b_0 s_0 + b_i s_i & a_0s_0+ a_i s_i \\ \bar{a}_0 s_0+\bar{a}_i s_i & \bar{b}_0 s_0 + \bar{b}_i s_i \end{array}\right),
\eeqn
with the Pauli matrices $s_i$ with $i=x,y,z$ and identity matrix $s_0$ acting on the spin space, and $a_{0(i)} (b_{0(i)})$ the $N\times N$ matrix. Here the reflection matrix is written in the basis $(c_{\uparrow},c_{\downarrow},c^{\dagger}_{\downarrow},-c^{\dagger}_{\uparrow})^{\rm T}$ with $c$ the electron annihilation operator. The blocks $a_0 s_0$ and $a_i s_i$ describing the Andreev reflection in the spin-singlet and spin-triplet channels respectively\cite{Stefanakis2002,Burset2014,Ebisu2016}.

Due to the scattering at NM/SC interface, the reflection matrix is determined by the superconductor as well as its coupling to the lead and take the form \cite{Nilsson2008}
\beqn
R(E)= 1-2\pi i W^{\dagger} \frac{1}{E-H_{\rm sc}+i \pi W^{\dagger}W}W,
\eeqn
where $H_{\rm sc}$ is the superconducting Hamiltonian and $W$ is the effective transverse modes coupling between the lead and the superconductor. If the superconductor also has time-reversal symmetry, it is straightforward to show that
\beqn
\hat{T} R(E) \hat{T}^{-1} = 1+2\pi i W^{\dagger} \frac{1}{E-H_{\rm sc}-i\pi W^{\dagger}W}W=R^{\dagger}(E), \nonumber
\eeqn
leading to the constrains of $\bar{a}_0=a_0^{\rm T}$, $\bar{a}_i=-a_i^{\rm T}$, $\bar b_0=b_{0}^{T}$ and $\bar b_i = -b_{i}^{\rm T}$. This means that in the spin-triplet Andreev reflection channel, there is a $\pi$ phase difference between the Cooper pairs entering and leaving CTSC as shown in Fig.~\ref{junctions}(a), while this $\pi$ phase difference is absent in the spin-singlet channel. If the superconductor also has inversion symmetry, namely being centrosymmetric, the effective coupling $W$ is spin-independent and the spin-triplet pairing and spin-singlet pairing are not mixed. In this case, the spin-triplet (spin-singlet) superconducting gap function can only induce Andreev reflection in spin-triplet (spin-singlet) channels. Therefore the observation of this $\pi$ phase in the centrosymmetric and time-reversal-invariant superconductor can serve as the definitive signal for spin-triplet parings. 

\begin{figure}
\centering
\includegraphics[width=1.\columnwidth]{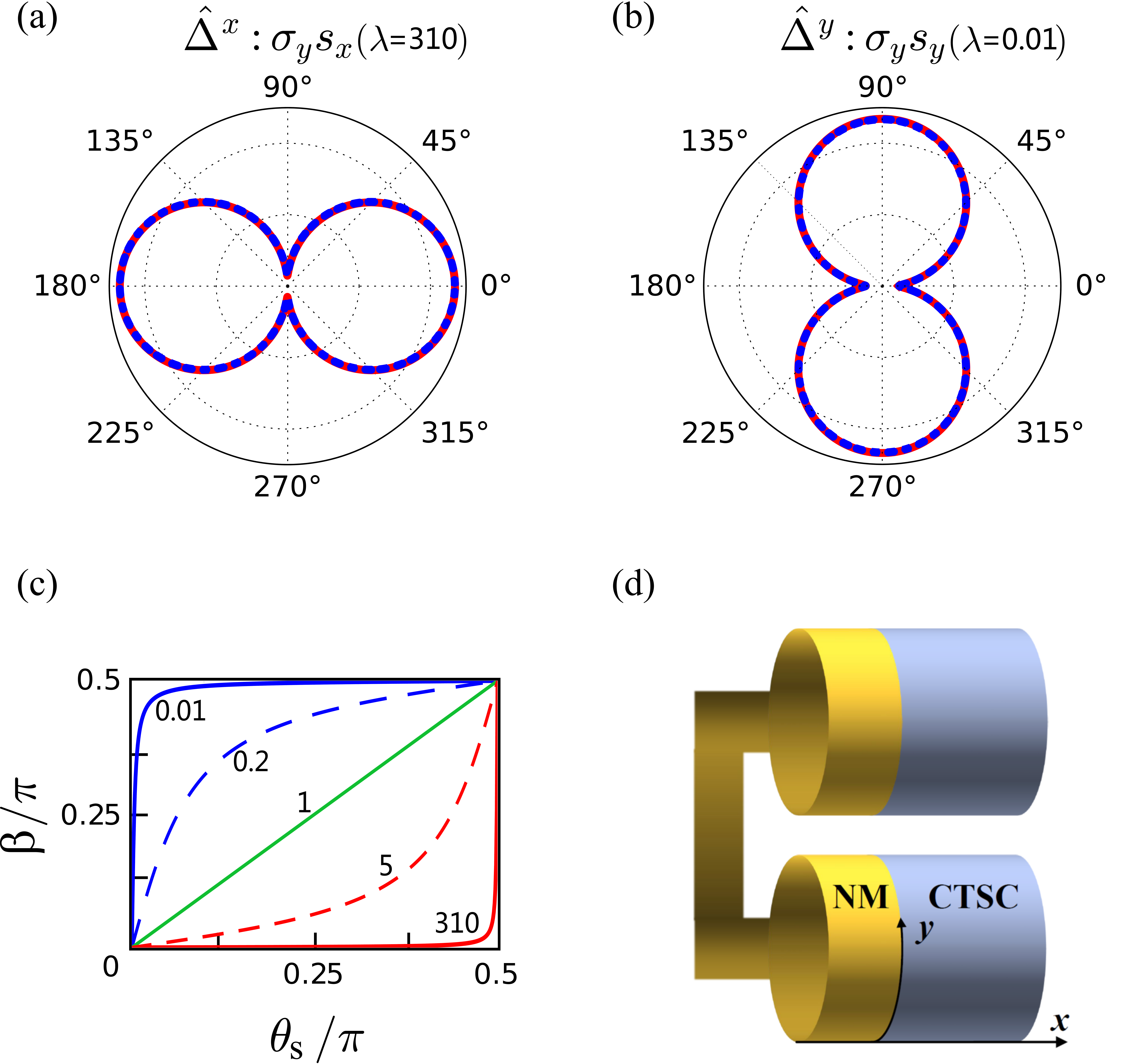}
\caption{The angle dependence of the anisotropic superconducting gap over the $k_z = 0$ Fermi contour for (a) $\Delta^x$ and (b) $\Delta^y$ respectively. The red line corresponds to the numerical result based on tight-binding models. Full gaps are robust and model-independent. (c) orbital-phase $\beta$ as a function of $\theta_{\rm s}$ for $\lambda$ from $0.01$ to $310$. (d) Josephson U-shaped junction with periodic boundary condition.}
\label{gap function}
\end{figure}

\subsection{Gap functions of M$_x$Bi$_2$Se$_3$}
Although the existence of the Andreev $\pi$-phase is independent of the specific form of the spin-triplet gap functions, to give a concrete example, we focus on M$_x$Bi$_2$Se$_3$, which is suggested to be a CTSC with spin-triplet parings. We consider the typical low energy effective Hamiltonian of M$_x$Bi$_2$Se$_3$\cite{Fu2010} 
\begin{eqnarray}\label{Ham-1}
H(k) &=& (h_0-\mu_s)\tau_z +{\Delta}^{i} \tau_x,
\end{eqnarray}
with $h_0$ the electron Hamiltonian which in the continue limit has the form \cite{Liu2010}
\begin{eqnarray}\label{Ham-e}
h_0 &=& m(k) \sigma_z  +v_z k_z\sigma_y+A(k_x s_y-k_y s_x)\sigma_x \nn
&&+R_1(k_{x}^{3}-3k_{x}k_{y}^{2})\sigma_xs_z+R_{2}(3k_{x}^{2}k_{y}-k_{y}^{3})\sigma_y,
\end{eqnarray}
in the basis $(c_{1\uparrow },c_{1\downarrow},c_{2\uparrow },c_{2\downarrow})^{\rm T}$ with $1,2$ indicating the conduction and valence band degrees of freedom, $\tau$, $\sigma$ and $s$ the Pauli matrices acting on Nambu, band and spin basis respectively, $\mu_s$ the chemical potential, $m(k)=m_0+m_1k_z^2+m_2(k_x^2+k_y^2)$, $R_1$ and $R_2$ the warping terms. ${\Delta}^{i}=\Delta_0\sigma_y s_i$ is the pairing potential and $i=x,y,z$ corresponding to three $s$-wave spin-triplet pairings. The possible spin-triple pairings includes $\Delta^z$ belonging to $A_{\rm 1u}$ representation and $\{\Delta^x, \Delta^y\}$ in the two-dimensional $E_u$ representation of D$_{3d}$ crystallographic point group\cite{Fu2014}. Note that $\Delta^x$ and $\Delta^y$ pairings can spontaneously break the continuous rotational symmetry down to two-fold, and the warping term $R_1$ and $R_2$ are essential for the presence of a full gap for ${\Delta}^y$ and ${\Delta}^x$ gap functions respectively\cite{Fu2014,Hao2017}. The rotational symmetry breaking observed in nuclear magnetic resonance\cite{Matano2016} and scanning tunneling microscopy measurements\cite{Tao2018}suggest the existence of $\Delta^x$ or $\Delta^y$ pairing. To have an intuitive understanding of $\{\Delta^x, \Delta^y\}$ pairing function, according to the Hamiltonian in Eq.~\eqref{Ham-1}, we plot the gap magnitudes as a function of the polar angle at $k_x$-$k_y$ plane given $k_z=0$ (red solid curves), in Fig.~\ref{gap function}(a)-(b) respectively. In the rest of this work, our numerical studies are built on the tight-binding model Hamiltonian Eq.~(\ref{Ham-1},\ref{Ham-e}) with its details shown in Appendix \ref{AP-D}. 
 
As the Fermi level lies in the conduction band, to simplify the analytical studies, we project the Hamiltonian in Eq.~\eqref{Ham-1} to the conduction band according to the quasi-degenerate perturbation theory\cite{Lowdin1951} (L\"{o}wding partitioning). After this projection, the Hamiltonian around the Fermi level has the form
\beqn\label{Ham-2}
H(k)=(\epsilon(k)-\mu_{\rm s})\tau_z + \hat{\Delta}^{i}(k) \tau_x,
\eeqn
where $\epsilon(k)= \hbar^2 (k_{\parallel}^2+k_{z}^2)/2m$, with $\parallel$ indicating the direction parallel to the $x$-$y$ plane and $m$ the effective mass. The three $s$-wave spin-triplet pairings in the two-band model \cite{Fu2010,Fu2014} are transformed to the typical $p$-wave pairings as
\beqn\label{Ham-3}
&&\hat{\Delta}^x \approx \tilde{A} k_x s_z+(\tilde{v}_z k_z+\tilde{R}_2k_y) s_x,\nn
&&\hat{\Delta}^y\approx\tilde{v}_z k_z s_y+\tilde{A} k_ys_z+\tilde{R}_1 k_x s_y, \nn
&&\hat{\Delta}^z \approx -\tilde{A}(k_xs_x+k_ys_y)+\tilde{v}_z k_z s_z, \nn
&&\left(\tilde{A},\tilde{R}_1\right)=\left(A,3R_1k_{\parallel}^2/\sqrt{2}\right)\Delta_0\sigma_1 \nn
&&\left(\tilde{v}_z,\tilde{R}_2\right)=\left(v_z,3R_2k_{\parallel}^2/\sqrt{2}\right)\Delta_0\sigma_2,
\eeqn
with $\sigma_1=\frac{\sigma_0}{m_{\rm 0}}+\frac{\sigma_z}{\mu_s}$ and $\sigma_2=\frac{\sigma_z}{m_{\rm 0}}+\frac{\sigma_0}{\mu_s}$ and $\Delta_0$ characterizing the gap magnitude. To simplify the analytical calculation, we replace the term $3k_x^2-k_y^2$ by its root mean square. According to Eq.~\eqref{Ham-3}, we plot the ${\Delta}^y$ and ${\Delta}^x$ gap magnitudes as a function of the polar angle in $x$-$y$ plane given $k_z=0$ in Fig.~\ref{gap function}(a)-(b) (dashed blue curves) respectively, which turns out to be a valid approximation for the gap functions obtained from Eq.~\eqref{Ham-1}. The gap functions in Eq.~\eqref{Ham-3} is consistent with the previous studies using eigenfunction projection methods \cite{Yip2013}. In this work, our analytical studies are based on the Eq.~(\ref{Ham-2},\ref{Ham-3}).

\subsection{Andreev reflection at NM/M$_x$Bi$_2$Se$_3$ interface}

The layer structure of the M$_x$Bi$_2$Se$_3$ makes the intra-layer transport more conductive than inter-layer transport. Therefore, we take the normal direction of the NM/M$_x$Bi$_2$Se$_3$ interface is along the $x$ direction. The Andreev reflection in the NM/M$_x$Bi$_2$Se$_3$ interface can be studied firstly in the typical $x$-$y$ and $x$-$z$ plane. In either plane containing $x$ direction, the Hamiltonian of NM/M$_x$Bi$_2$Se$_3$ junction is simplified to a two-dimensional form
\beqn\label{Ham-4}
H_{J}=\Theta(-x)H_{\rm NM}+ V \delta(x)+\Theta(x) H_{\rm sc},
\eeqn
where 
\beqn
H_{\rm NM}=\left(\frac{\hbar^2( -\partial^2_x+k_2^2)}{2m}-\mu_{\rm n}\right)\tau_z \nonumber
\eeqn
is the Hamiltonian for the normal metal, with $k_{2}$ the in-plane momentum perpendicular to the $x$ direction and $\mu_{\rm n}$ the chemical potential. $V$ is the height of $\delta$-shaped barrier at the interface. The superconducting Hamiltonian $H_{\rm sc}$ generally takes the form
\beqn\label{Ham-p}
H_{\rm sc}=(-\frac{ \hbar^2 \partial_x^2}{2m_{x}}+\frac{\hbar^2 k_2^2}{2m_2}-\mu_{\rm s})\tau_z  + \Delta \tau_x ( \lambda i\partial_x s_i +  k_2 s_j),\nonumber
\eeqn
with $s_{i}$ the Pauli matrix acting on spin space. The parameters $\Delta$ and $\lambda$ reflect the amplitude and the shape of the gap function. Specific forms of $\Delta^{z,x,y}$ for $x$-$y$ and $x$-$z$ planes are listed in Table.~\ref{table:table1}. For $\hat{\Delta}^z$ gap, it corresponds to $\lambda =1$ in $x$-$y$ plane and $\lambda >1$ in $x$-$z$ plane. For $\hat{\Delta}^{x(y)}$ gap, it always corresponds to $\lambda >(<)1$, giving a dumbbell-shaped gap with major axis along (perpendicular to) the $x$ direction, in either $x$-$y$ or $x$-$z$ plane. 

\begin{table}[t]
\begin{center}
\caption{The parameters for three odd-parity gap functions in Eq.~\eqref{Ham-3}.}
\label{table:table1}
\begin{adjustbox}{max width=0.9\textwidth}
\begin{tabular}{|c|c|c|}
  \hline
  \multirow{2}{*}{\diagbox{gap}{plane}}&
  x-y plane & x-z plane \\
   & $k_2=k_y$, $m_2=m_{\parallel}$ & $k_2=k_z$, $m_2=m_{z}$ \\
    \hline
   $\Delta^z$&  $\lambda=1$, $s_i=s_x$, $s_j=s_y$ & $\lambda>1$, $s_i=s_x$, $ s_j=s_z$ \\
   \hline
  $\Delta^x$ & $\lambda>1$, $s_i=s_z$, $s_j=s_x$ & $\lambda>1$, $s_i=s_z,s_j=s_x$ \\
  \hline
   $\Delta^y$ &$\lambda<1$, $s_i=s_y$, $s_j=s_z$ & $\lambda<1$, $s_i=s_y$, $s_j=s_y$ \\
  \hline
\end{tabular}
\end{adjustbox}
\end{center}
\end{table}

\begin{figure}
\centering
\includegraphics[width=1.0\columnwidth]{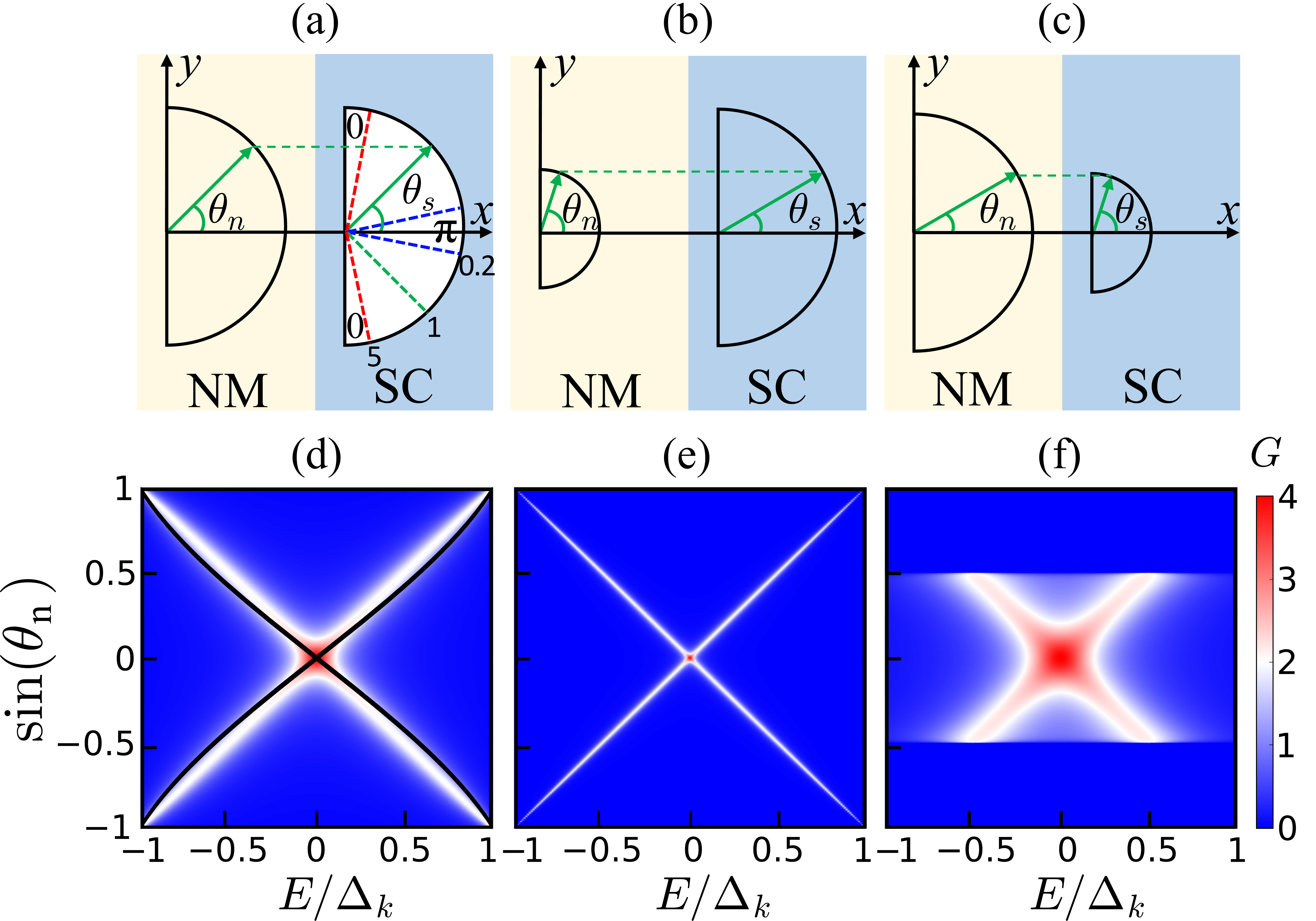}
\caption{(a) Fermi surface contours of both NM and SC regions for $r=1$. In this case, $\theta_{\rm s}=\theta_{\rm n}$ and transition angle $\theta_{\rm c}=\tan^{-1} \lambda$. The blue, green and red dashed lines indicate transition angle for $\lambda=0.2$, $1$ and $5$, respectively. (b) Fermi surface contours for $r>1$. (c) Fermi surface contours for $r<1$. (d)-(f) Conductance of a NM/CTSC junction for $r$ (d) $=1$, (e) $>1$ and (f) $<1$, respectively. A finite interface barrier is fixed, $Z=2$. The black line in (d) represents the dispersion of edge states.}
\label{angle-dependent-r}
\end{figure}

Considering the reflection in $x$-$y$ plane, the three gap functions can be written as $\Delta (\lambda k_x s_{i} + k_y s_{j})$ with $\lambda=1$, $\lambda>1$ and $\lambda <1$ corresponding to $\Delta^x$, $\Delta^y$ and $\Delta^z$ respectively. To simplify the analytical discussion, we take the periodic boundary condition along the $y$ direction for the two superconductors so that the two-dimensional junction can be simplified as the combination of numerous one-dimensional (1D) Josephson junctions with different $k_y$ as shown in Fig.~\ref{gap function}(d). We denote the magnitude of the Fermi momentum of normal (superconducting) regions by $k_{f\rm n(s)}$. $\theta_{\rm s}$ is always associated with $\theta_{\rm n}$ through the conservation of momentum $k_y$. Therefore, for a given channel with $k_y$, we have
\begin{gather}
k_y=k_{f\rm n}\sin\theta_{\rm n}=k_{f\rm s}\sin\theta_{\rm s}.
\end{gather}
Note that if both $m_{x}=m_2=m$ and the chemical potential $\mu_{\rm s}=\mu_{\rm n}$, the Fermi momentum $k_{f\rm n}$ and $k_{f\rm s}$, and thus $\theta_{\rm n}$ and $\theta_{\rm s}$, are equal as shown in Fig.~\ref{angle-dependent-r}(a). Otherwise, the Fermi momentums are generally different, cause $r\neq1$ and $\theta_{\rm n} \neq \theta_{\rm s}$ as shown in Fig.~\ref{angle-dependent-r}(b,c). 

Given the incident angle $\theta_{\rm n}$ and associated $\theta_{\rm s}$, the reflection matrix at the interface can be obtained as \cite{Tanaka1995,Yokoyama2007}
\beqn\label{RM-1}
R^{\pm}&=&\frac{\left(\begin{array}{cc} -b \cos(\alpha \mp \beta)  & -2\tilde{r} \\ 2\tilde{r}  & -b^{*} \cos(\alpha \mp \beta) \end{array}\right)}{\left(\tilde{r}^2+4\tilde{Z}^2+1\right) \cos (\alpha \mp \beta )+2 i \tilde{r} \sin (\alpha \mp \beta )},
\eeqn
\beqn
&&b=\tilde{r}^2+(2 \tilde{Z}+i)^2, \ \ \tilde{r}=r\cos\theta_{\rm s}/\cos\theta_{\rm n}, \ \ \tilde{Z}=Z/\cos\theta_{\rm n}, \nonumber 
\eeqn
with $R^{\pm}$ the scattering matrix in the spin-channel parallel and anti-parallel with $z$ direction (see Appendix \ref{AP-A} for derivation of this result), $Z=V /\hbar v_{f\rm n}$ the dimensionless height of the $\delta$-shaped barrier potential, $r=v_{f\rm s} /v_{f\rm n}$ the mismatch between Fermi velocity of superconducting and normal regions, $\alpha=\cos^{-1}(E/\Delta_{\bm k})$ with $\Delta_{\bm k}=\Delta \sqrt{\lambda^2 k_{x}^2+k_y^2}$ the gap amplitude and
\beqn\label{orbital-phase}
\beta=\tan^{-1} (k_y/(\lambda k_{x \rm s}))=\tan^{-1} (d_j/d_i),
\eeqn 
which characters the $d$-vector direction. We plot in Fig.~\ref{gap function}(c) $\beta$ as a functions of $\theta_{\rm s}$ for different $\lambda$.

Although the reflection matrix looks very complicated, we found two properties that are independent of dimensionless $r$ and $Z$. The first is the $\pi$ phase difference between the two Andreev processes characterized by the upper and lower off-diagonal block of the reflection matrix in Eq.~\eqref{RM-1}, which is consistent with our general analysis in Eq.~\eqref{R-matrix}. The second is related to the phase factor $\beta$ which can result in the perfect Andreev reflection through the condition $\cos(\alpha \pm \beta)=0$. This further leads to the conductance peak at $E=\pm \Delta_{\bm k} \sin(\beta)$, which has topological origin and actually comes from the edge states induced resonant scattering \cite{Liu2015a,Manikandan2017,Neha2019,Kim2020}. We plot in Fig.~\ref{angle-dependent-r}(d-f) the conductance as function of $\theta_{\rm n}$ and incident energy of particles for $\lambda=1$. The connection between the orbital-phase caused perfect Andreev reflection and the topological edge states can be shown more straightforwardly for $r=1$. In this case, the energy dispersion of the helical edge states is calculated, based on Eq.~\eqref{Ham-3}, as function of $\theta_{\rm n}$ (black curves in Fig.~\ref{angle-dependent-r}(d)), which matches the resonance peak at $E=\pm \Delta_{\bm k} \sin(\beta)$ very well. As the phase $\beta$ is determined by the $d$-vector, we dubbed it orbital-phase.

\section{Josephson effect in the CTSC/NM/CTSC junction}
In this section, we study the Josephson effect, based on the Andreev bound states, in the CTSC/NM/CTSC junction, and show that the competition between the intrinsic $\pi$ phase and the orbital-phase $\beta$ can result in a Josephson $\pi$-junction. 

\begin{figure}[t]
\centering
\includegraphics[width=1\columnwidth]{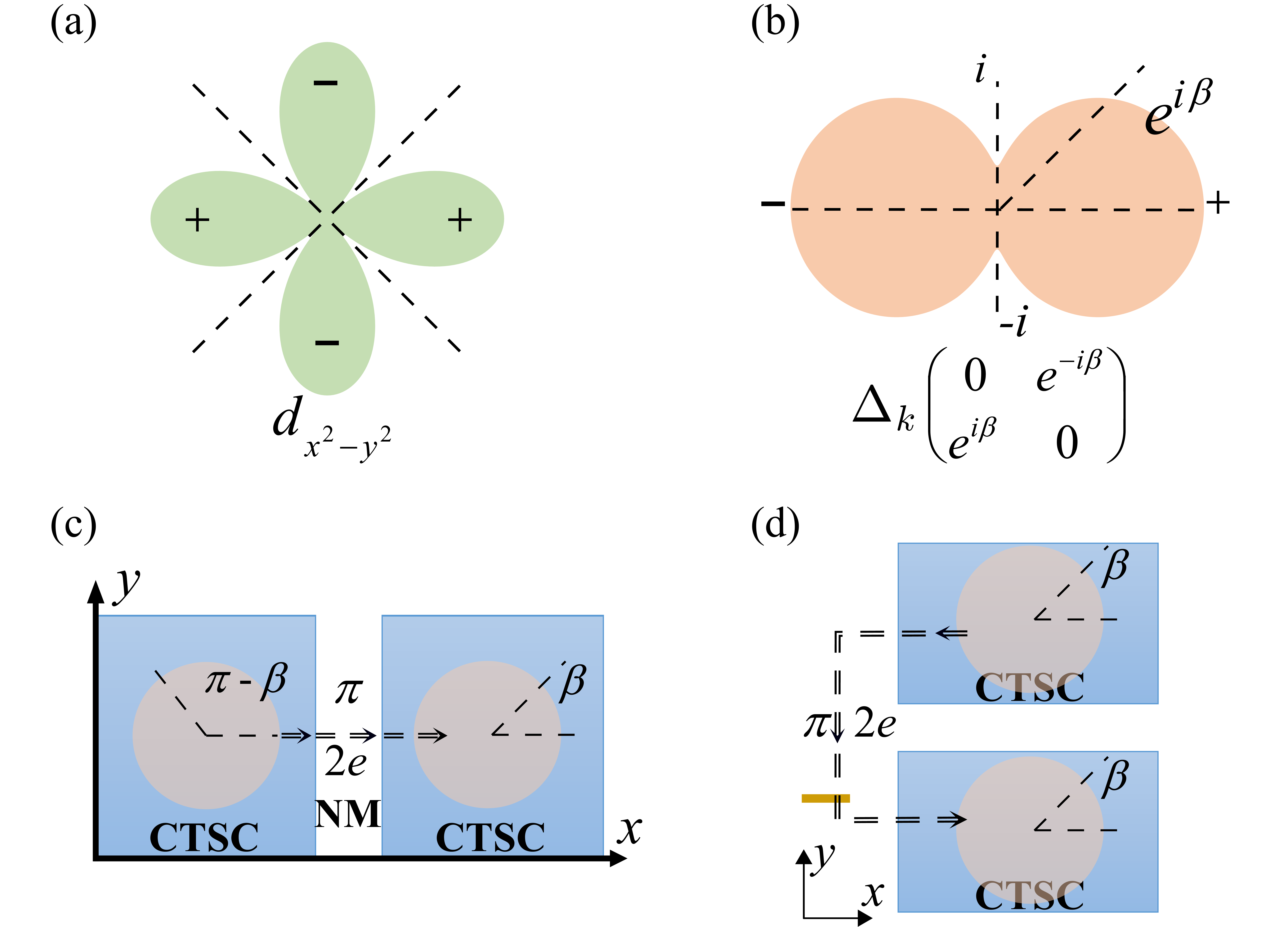}
\caption{(a) The angle dependence of the superconducting gap over the $k_z= 0$ Fermi contour for $d$-wave superconductor. $\pm$ is the sign of the gap amplitude. (b) The angle dependence of the superconducting gap for $\lambda=5$ with an additional orbital-phase $\beta$. (c) (d) Processes of Cooper pair through conventional and U-shaped Josephson junction, the circle graphs are the superconducting gaps for $\lambda=1$.}
\label{orbital phase}
\end{figure}

\subsection{An intuitive pictures}\label{3-A}
Here, we provide an intuitive discussion of how the various phases affect the CPR of the Josephson effect. To avoid possible confusion, we first take the flux induced phase $\phi=0$ in this subsection. In this case, the spin-singlet gap function can only take the positive or negative sign of its gap amplitude. Taking $d$-wave superconductor as an example, due to the $d$-orbital symmetry, the sign of the gap amplitude alternate every $\pi/2$ (Fig.~\ref{orbital phase}(a)), which means the orbital-phase in spin-singlet order parameters can only take 0 or $\pi$. The situation is more complicated for spin-triplet gaps because the gap amplitudes are described by a vector which may result in a continuously relative phase among the spin-triplet pairings. Generally, the gap function
\beqn
\Delta(\lambda k_{x \rm s} s_x + k_y s_y) = \Delta_{\bm k} \left(\begin{array}{cc} 0 & e^{i\beta} \\ e^{-i\beta} & 0 \end{array}\right)
\eeqn
implies that the pairing states $|\uparrow \uparrow\rangle$ and $|\downarrow \downarrow\rangle$ have the relative phase $\beta$(Fig.~\ref{orbital phase}(b)) which can varies continuously in the range $[0,2\pi)$. Note that this phase result in the resonance conductance peak and thus is not a conceptual definition. These phases can be brought to the Josephson junction the Andreev reflections which split or recombine Cooper pairs on the NM/SC interface. Note that the Cooper pair transportation through a Josephson junction goes through three processes: the Cooper pair being split into two electrons in the SC/NM interface, then passing through the normal region, and finally being recombined into a Cooper pair at another NM/SC interface. Each process may contribute to the phases, which finally affect the CPR of the Josephson effect. For example, when the spin degeneracy is lifted in the normal region by the Zeeman effect or spin-orbit coupling \cite{Pershoguba2015,Pientka2017,Cayao2018,Ren2019}, the electron and hole transmission through the junction can accumulate unusual phases, causing various interesting anomalous Josephson effects \cite{Ryazanov2001,Champel2008,Lu2018,Zhang2018a}. However, this is not the main purpose of this work and we can ignore these phases by adopting both time-reversal and inversion symmetry in the junction region. Here, we focus on the phase effect through the split and recombination of the spin-triplet Cooper pairs, which are implemented through Andreev reflection on the NM/CTSC interface.

When the Josephson junction adopts the conventional geometry structure (Fig.~\ref{junctions}(b)), the orbital-phase in these two processes with given $k_y$ contributes to a $\pi$ phase as schematically shown in Fig.~\ref{orbital phase}(c). This $\pi$ phase is completely offset by the Andreev $\pi$-phase so that the CPR is only determined by the flux phase $\phi$ which gives rise to the CPR very similar with that for the $s$-wave Josephson junction (see Appendix \ref{AP-B} for details). In the U-shaped junction(Fig.~\ref{orbital phase}(d)), the orbital-phase in each channel with given $k_y$ contributes to a $2\beta$ phase which combines with the intrinsic $\pi$ phase leads to a $\pi$-junction and $0$-junction for the incident angle $0$ and $\pm \pi/2$ respectively, which is very different from the conventional $s$-wave Josephson junction. Below, we focus on the U-shaped junction and show details on how to use it to distinguish the spin-triplet pairings.

\begin{figure}[t]
\centering
\includegraphics[width=1\columnwidth]{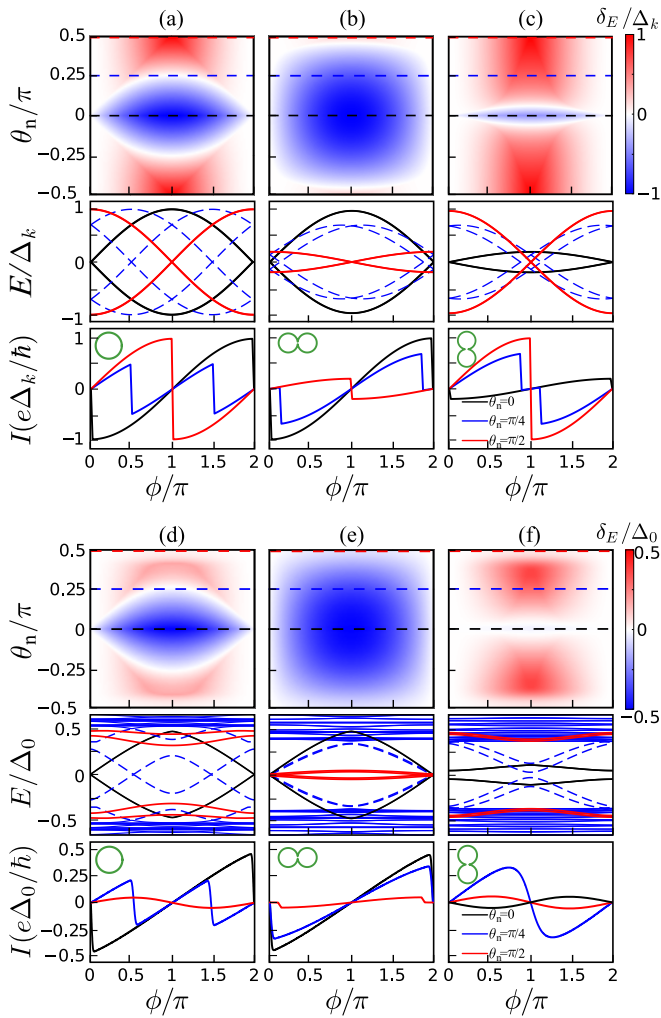}
\caption{The transition between $0$ and $\pi$-junction in perfect contact cases. (a)-(c) Upper and middle pattern: The Andreev levels and CPRs corresponding to $\theta=\frac{\pi}{4}$(blue dashed line), $0$(black solid line) and $\frac{\pi}{2}$(red solid line).  Lower pattern: Normalized free energy difference $\delta_F$ derived from analytical calculations in 1D U-shaped CTSC/NM/CTSC junction corresponding to $\lambda=1$, $5$ and $0.2$, respectively. (d)-(f) Andreev levels, CPRs and $\delta_F$ obtained from the numerical calculations corresponding to ${\Delta}^z$, ${\Delta}^x$ and ${\Delta}^y$, respectively. The inset shows the angle dependence of the superconducting gap over the $k_z=0$ Fermi contour.}
\label{1D-ideal}
\end{figure}

\subsection{Perfect contact cases}

We first consider the case with perfect Fermi velocity match ($r=1$) and zero barrier potential ($Z=0$). In the limit of a short junction, we neglect the possible scattering in the junction and calculate the Andreev levels in each channel with different $\theta_n$ directly from solving the equation $\rm{Det}[I-R^{\pm}(\phi/2)R^{\pm}(-\phi/2)]=0$ \cite{Beenakker1991,Supp} as
\beqn\label{AL-2}
E=\pm\Delta_{\bm k}\cos(\frac{\phi-\pi }{2}\pm \beta).
\eeqn
Meanwhile, the associated Josephson current can be calculated through the standard form
\begin{eqnarray}\label{JC}
I=\frac{2e}{\hbar} \sum_{E_n<0}\frac{\partial E_n}{\partial \phi},
\end{eqnarray} 
at zero temperature. We plot in Fig.~\ref{1D-ideal}(a-c) the Andreev levels and CPRs for $\lambda = 1$, $5$ and $0.2$ respectively. The crucial orbital-phase $\beta$ is related to the incident angle by $\beta=\tan^{-1}(\tan\theta_{\rm n}/\lambda)$ for $r=1$. For normal incidence with $\beta=0$, the Andreev levels cross at $\phi=0$, causing the $\pi$-junction with a period of $4\pi$ for all $\lambda$. For $\theta_{\rm n}=\beta=\pm \pi/2$, the cross of Andreev levels is shifted to $\phi=\pi$ (red curves in Fig.~\ref{1D-ideal}(a-c)), resulting in the Josephson 0-junction. This implies that the transverse modes for low and high incident angles perform as Josephson $\pi$- and $0$-junction respectively. The low-angle incident channels and the high-angle incident channels contribute opposite Josephson currents as shown in Fig.~\ref{1D-ideal}. The transition angle between $\pi$- and $0$-junction is determined by the gap anisotropy which is characterized by $\lambda$. In short, the transition angle, $\theta_{\rm c}$, in the first quadrant is larger, equal and smaller than $\pi/4$ for $\lambda<1$, $\lambda=1$ and $\lambda>1$ respectively. To demonstrate this clearly, we define the free energy difference $\delta F(\phi)$ given the incident angle $\theta_n$ as 
\beqn\label{FE}
\delta F(\phi,\theta_n)= \sum_{E_n<0}\left(\frac{E_n(\phi,\theta_n)-E_n(\phi=0,\theta_n)}{N_E}\right),
\eeqn
where $N_E$ is the number of occupied ABSs. Note that $\delta F(\phi=\pi) >0$ and $\delta F(\phi=\pi)<0$ correspond to the 0-junction and $\pi$-junction respectively while $\delta F(\phi=\pi)=0$ gives the transition condition. For $\lambda=1$ (Fig.~\ref{1D-ideal}(a)), we plot $\delta F$ as a function of $\phi$ and $\theta_n$ which shows that $\delta F(\pi)=0$ happens at $\theta_n=\pm \pi/4$. In Fig.~\ref{1D-ideal}(b) and Fig.~\ref{1D-ideal}(c), we calculate $\delta F$ for $\lambda = 5$ and $\lambda=0.2$ which show that the transition angle $\theta_c$ is decreased and increased from $\pi/4$, respectively, in the first quadrant. Above analytical results perfectly match our numerical calculations, as shown in Fig.~\ref{1D-ideal}(d-f), based on 1D tight-binding model with interorbital and odd-parity pairing forms. (See Appendix \ref{AP-D} for the model.)

\begin{figure}
\centering
\includegraphics[width=1.0\columnwidth]{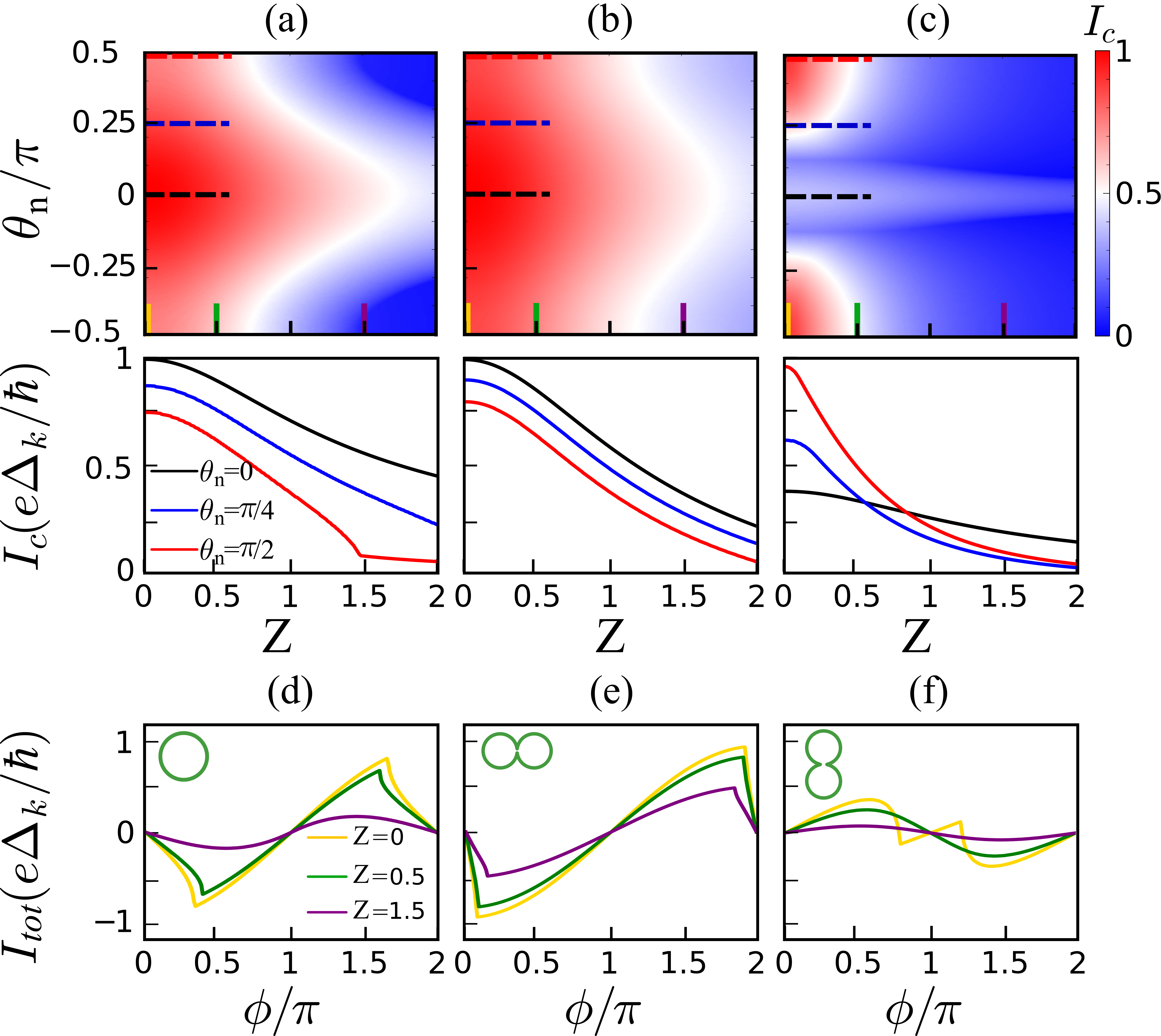}
\caption{The Josephson current in imperfect contact case with Fermi velocity mismatch $r=2$. (a-c) Upper pattern: $I_c$ as a function of $\theta_n$ and Z for $\lambda$=1, 5 and 0.2, respectively. Lower pattern: $I_c$ as a function of $\theta_n$ extracted from upper pattern. (d-f) $I_{\rm tot}$ as a function of $\phi$ for $\lambda$=1, 5 and 0.2, respectively. Different colored curves correspond to different $Z$, as the short lines in the top figures.}
\label{non-ideal}
\end{figure}

\subsection{Robustness of $\pi$-junction against imperfect NM/CTSC interface} 

In this section, we consider how the imperfect factors including the velocity mismatch and the finite barrier potential, which normally inevitably exist in Josephson junctions, affect the results \cite{Asano2006,Sawa2007,Yokoyama2007a}. In the short junction limit, we can analytically obtain the Andreev levels with given incident angle $\theta_n$ (Fig.~\ref{angle-dependent-r}(a-c)) as
\beqn\label{AL-1}
\frac{E}{\Delta_{\bm k}} &=& \mp \left[\sqrt{\tilde{T}_{\rm n}} \cos(\frac{\phi - \pi}{2})\cos\beta \pm \sqrt{1-\tilde{T}_{\rm n} \sin^2 \frac{\phi}{2}}\sin\beta\right], \notag \\
\
\eeqn
with $\tilde{T}_{\rm n}=\tilde{r}^2/(\tilde{r}^2+4 \tilde{Z}^2)$ the transmission coefficient when the superconductor becomes normal metal. Usually, the imperfect factors reduce transparency and make the $\tilde{T}_{\rm n}<1$. In this case, the effect of orbital-phase $\beta$ on Andreev levels can no longer be simply attributed to phase shift as in Eq.~\eqref{AL-2}. Meanwhile, it is noteworthy that the first and second terms on the right side of the Eq.~\eqref{AL-1} have the same form as the Andreev levels of the one-dimensional topological and conventional Josephson junction respectively, while their ratio is tunned by the orbital-phase $\beta$. This can be seen clearly in the case of $r=1$. At $\theta_{\rm n}=0$ corresponds to $\beta=0$, Andreev levels in Eq.~\eqref{AL-1} is a topological $\pi$-junction with a period of $4\pi$\cite{Fu2009,Peng2016,Haim2019,Razmadze2020}. As the incident angle increases, particularly at $ \theta _ {\rm n} = \pm \pi/2 $ corresponds to $\beta = \pm \pi/2$, the CPR shows a conventional 0-junction with a period of $2 \pi$. As a result, it is still true that the low-angle incident channels and the high-angle incident channels contribute opposite Josephson currents. Besides, given the Fermi mismatch $r$ and the dimensionless barrier potential $Z$, the normal transmission coefficient $\tilde{T}_n$ decrease with increasing $|\theta _ {\rm n}|$ from 0 to $\pi/2$. Therefore the critical current $I_{\rm c}$ of the high incident-angle channel is suppressed more by the imperfect factors.

In Fig.~\ref{non-ideal}(a-c) and taking the Fermi velocity mismatch $r=2$, we plot $I_{\rm c}$ as a function of the incident angle $\theta_{\rm n}$ and $Z$ for $\lambda=1$, $\lambda=5$ and $\lambda=0.2$ respectively. In Fig.~\ref{non-ideal}(a-b), $I_c$ in the low-angle incident channel is always larger than that in the high-angle incident channel. In Fig.~\ref{non-ideal}(c) with $\lambda=0.2$, as the gap $\Delta_k$ is larger in high incident angle, the high-angle incident channel has larger critical current. These are consistent with our analytical results according to Eq.~\eqref{AL-1}. Thus for $\lambda \ge 1$, the net Josephson current, integrated over the incident angle $I_{\rm tot}=\int I d\theta_n$ is always dominated by the CPRs of low-angle incident channels which give the Josephson $\pi$-junction (Fig.~\ref{non-ideal}(d-e)). On the contrary, for $\lambda \ll 1$, the net current is dominated by the CPRs of high-angle incident channels which give the Josephson $0$-junction (Fig.~\ref{non-ideal}(f)). We can get similar results for $r=1$ and $r<1$ which are shown in Appendix \ref{AP-C}. Therefore, we conclude that the imperfect factors such as the Fermi velocity mismatch and interface barrier make CRP more inclined to be $\pi$-junction.

\begin{table}[b]
\begin{center}
\caption{The phenomenon of CPRs for three odd parity gap functions in various types of JJs. The normal region of $\rm U_{x(y)}$-shaped junction contacts the same side of the superconductors and the normal direction of both NM/SC interfaces is along $x(y)$ direction(Fig.~\ref{JC-3D}). The normal region of conventional junction, in turn, contacts to opposite side of superconductors(Fig.~\ref{N-junction}).}
\label{table:table2}
\begin{adjustbox}{max width=1\textwidth}
\begin{tabular}{|c|c|c|c|}
  \hline
  \multirow{2}{*}{\diagbox{gap}{type}}&$\rm U_{x}$-shaped  & $\rm U_{y}$-shaped &  Conventional\\
&x & y & x/y\\
    \hline
   $\Delta^z(\lambda=1)$&  $\pi$-junction &  $\pi$-junction &   $0$-junction\\
   \hline
  $\Delta^x(\lambda>1)$ &   $\pi$-junction &  $0$-junction &   $0$-junction\\
  \hline
   $\Delta^y(\lambda<1)$ &   $0$-junction &   $\pi$-junction &   $0$-junction \\
  \hline
\end{tabular}
\end{adjustbox}
\end{center}
\end{table}

\begin{figure}
\centering
\includegraphics[width=1.\columnwidth]{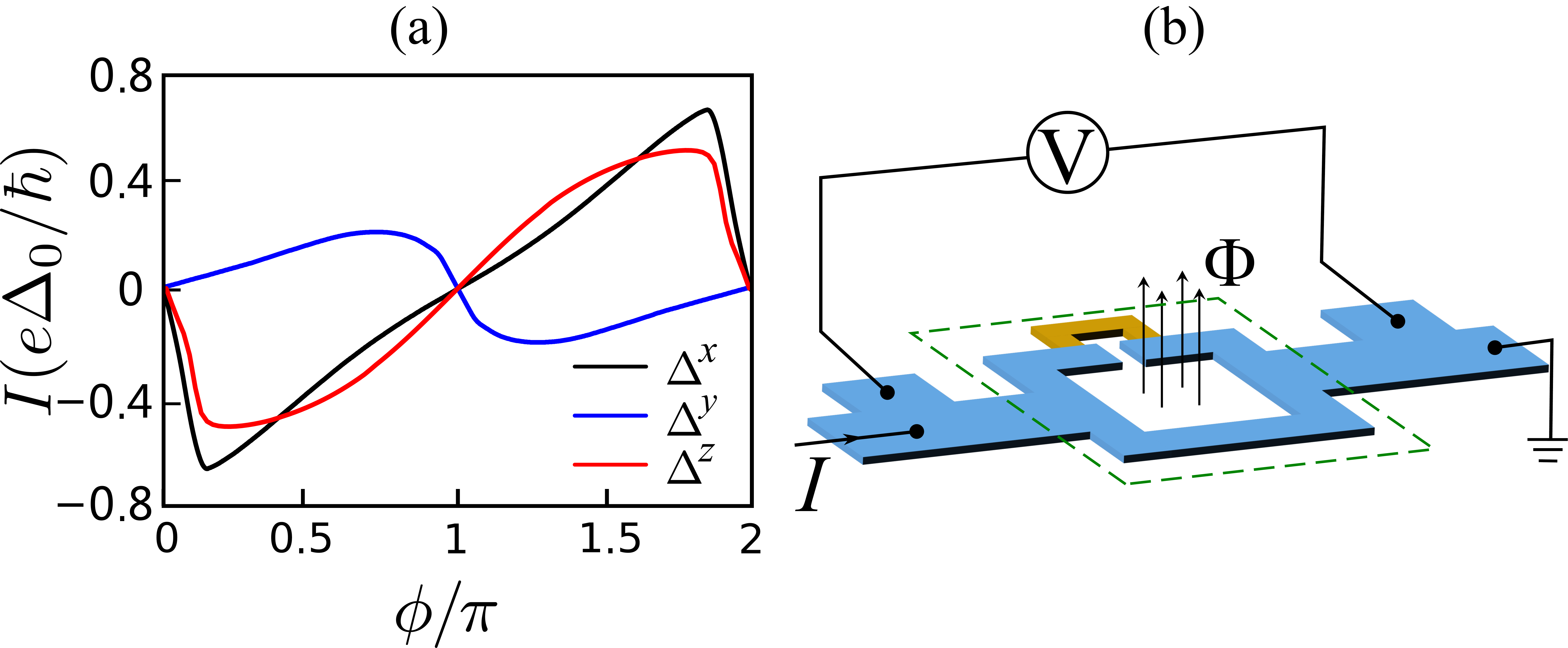}
\caption{(a) The Josephson currents for different pairings, showing the $\pi$-junction for the $\hat{\Delta}^x$ and $\hat{\Delta}^z$ pairings and $\pi$-junctions for the $\hat{\Delta}^y$ pairing. (b) Schematic of the loop device of magnetoresistance experiment. $\Phi$ is the applied magnetic flux.}
\label{JC-3D}
\end{figure}

\section{Josephson $\pi$-junction in M$_x$Bi$_2$Se$_3$}
Now, let us apply our analysis on the M$_x$Bi$_2$Se$_3$/NM/M$_x$Bi$_2$Se$_3$ junction. Considering the transport in the $x$-$y$ plane with the normal of the interface along the $x$ direction, the gap functions $\hat{\Delta}^{z}$, $\hat{\Delta}^{x}$ and $\hat{\Delta}^y$ correspond to $\lambda=1$, $\lambda>1$ and $\lambda<1$ so as to give the CPR of Josephson $\pi$-, $\pi$- and $0$-junction, respectively. According to the experimental parameters\cite{Liu2010}, $\lambda$ is much larger than 1 for $\hat{\Delta}^{x}$, ensuring a stable $\pi$-junction due to the larger transition angle. 
Similarly, for the normal direction of the interface along $y$ direction, the CPRs of junction with gap function $\hat{\Delta}^{z}$ and $\hat{\Delta}^{y}$ should provide stable $\pi$-junction, and $\hat{\Delta}^{x}$ provides $0$-junction. 

To further verify our results, we perform the numerical calculation using a three-dimensional tight-binding model based on Eq.~\ref{Ham-1} and plot the Josephson current versus the phase difference $\phi$ in Fig.~\ref{JC-3D}. We found that for $\hat{\Delta}^{x}$ and $\hat{\Delta}^{z}$ pairings, the CPR indicates the Josephson $\pi$-junction, while it gives the Josephson 0-junction for $\hat{\Delta}^{y,}$ pairings. Thus the numerical results are a perfect match for our analysis. For the junction along the $y$ direction, the results are the same as long as we exchange $x$ and $y$ index in the gap functions. Meanwhile, in the conventional Josephson junction geometry with the normal lead contacting to the opposite direction of the superconductors, all of the gap functions give the $0$-junction (see Appendix \ref{AP-B} for the details), which is consistent with our analysis in section \ref{3-A}. In Table.~\ref{table:table2}, We summarized the CPR for the three gap functions, which shows the different CPRs for the U-shaped junctions with various gap functions. Accordingly, we conclude that the U-shaped CTSC/NM/CTSC junction can be utilized to detect the spin-triplet gap functions while it may fail in the conventional junction geometry.

\section{Experimental realization and discussion}
It is well known that when a superconducting loop contains odd number of Josephson $\pi$-junction, it will create spontaneously a half-quantum magnetic flux (HQF)\cite{Geshkenbein1987,Sigrist1992,Tsuei1994}. The HQF could be directly observation observed directly by a magnetoresistance oscillation measurement based on the Little-Parks effect. The Little-Parks effect demonstrated the oscillations of the superconducting transition temperature $T_c$, and hence the sample resistance in the transition regime, as a function of the applied magnetic field resulting from the corresponding variation of the free energy\cite{Little1962}. The minimum of the free energy, and the minimum of resistance, is achieved when applied magnetic flux become $(n+1/2)\Phi_0$ for a superconducting loop contains odd number of $\pi$-junctions, instead of $n\Phi_0$ for conventional $s$-wave superconductors, where $n$ is an integer number and $\Phi_0$ the magnetic flux quantum, $\Phi_0 = h/2e$, where $h$ is the Planck constant and $e$ the elemental charge\cite{Byers1961,Deaver1961}. Considering a M$_x$Bi$_2$Se$_3$ loop as shown in green dashed rectangle in Fig.~\ref{JC-3D}(b), which has a nanometer-sized fracture and alternative metal bridge attached to one side of loop, we expect to observe the spontaneous HQF if the gap function is odd-parity. For a loop with outer dimensions of 800 nm by 800 nm and line width of 100 nm, the oscillation period of magnetic field intensity $H\approx 42$ Oe. One can fabricate a square-shaped loop device using textured M$_x$Bi$_2$Se$_3$ thin films. Taking Cu$_x$Bi$_2$Se$_3$ as an example, it has superconducting transition temperature $T_c$ of $3.8$ K at $x=0.21$ and in-plane Ginzburg-Landau coherence length $\xi$ of $14$ nm \cite{Hor2010,Kriener2011}. Recent STM experiment\cite{Tao2018} of Cu$_x$Bi$_2$Se$_3$ ($x$=0.39) have observed larger coherence length of 30.83 nm along long axis even under $H=2000$ Oe. The resistance oscillations as a function applied magnetic field of the four terminal device, shown in Fig.~\ref{JC-3D}(b), should be observed at temperature near $T_c$, using a dc technique. The maximum of resistance at $H=42n$ Oe is the key signature of the formation of Josephson $\pi$-junction, and thus the spin-triplet parings of CTSC.

In conclusion, we studied the Andreev reflection and Josephson effect in the NM/CTSC hybrids. There are two important phases in the spin-triplet Andreev reflection channels. The first is the so-called Andreev $\pi$-reflection, telling the $\pi$-phase difference between the spin-triplet Coopers entering and leaving CTSC. The second is the orbital-phase, which originated from the phase difference between the spin-triplet pairing states with different spin polarization. In the U-shaped Josephson junction (Fig.~\ref{junctions}(b)), the competition between these two phases leads to the CPR of $\pi$-junction (0-junction) in the channel with low (high) incident angle. The net CPR of the Josephson junction, taking into account the contribution from all channels, depends on the anisotropy of the odd-parity gap functions. The imperfect contacts such as the Fermi velocity mismatch and the barrier potential always make the CPR inclined to be the $\pi$-junction which is helpful for the experimental observation. When the junction takes the traditional geometry structure, its normal region contacting the opposite sides of the superconductor (Fig.~\ref{junctions}(c)), the competition of these two phases in CPR completely disappears, which gives rise to the Josephson $0$-junction in all channels. We then conclude that the observation of the Josephson $\pi$-junction in the U-shaped junction can serve as a definitive signal for the odd-parity CTSC. At last, we provide a brief discussion of the experimental realization.

\section*{Acknowledgement}

We would like to thank Bo Lu, Zhi Gang Cheng and Chao-Xing Liu for fruitful discussions. X. Liu acknowledges the Ministry of Science and Technology of China (Grant No. 2016YFA0401003) and the National Natural Science Foundation of China (Grant No. 11674114).

\appendix
\section{Scattering matrix at the NM/CTSC interface}\label{AP-A}
In this section, we study the reflection matrix at the NM/CTSC interface described by the Hamiltonian in Eq.~\eqref{Ham-4}. The interface is perpendicular to $x$ direction and we assume that there are not scatterings take place in $y$ direction. Let us start by considering the case of an incident electron, incoming from the NM towards the interface
\beqn\label{A1}
\Psi^{I}=\left(\begin{array}{c} 1  \\0 \\0\\ 0 \end{array}\right)e^{i k_x x}f(y),
\eeqn
where $k_x$ is the wave number of the electron in NM. As a result,electron-like (e) and hole-like (h) quasi-particles with opposite spins $\uparrow(\downarrow)$ are transmitted into the superconductor:

\beqn\label{A2}
\Psi^{T}&=&t_{e\uparrow e\uparrow}\widetilde\psi_{e\uparrow}+t_{e\downarrow e\uparrow}\widetilde\psi_{e\downarrow}+t_{h\downarrow e\uparrow}\widetilde\psi_{h\downarrow}+t_{h\uparrow e\uparrow}\widetilde\psi_{h\uparrow}\nonumber\\
&&\widetilde\psi_{e\uparrow}=\left(\begin{array}{c} 1  \\0 \\0\\ e^{-i (\alpha-\beta)}\end{array}\right)e^{i \widetilde k_x x}f(y),\nonumber\\
&&\widetilde\psi_{e\downarrow}=\left(\begin{array}{c} 0  \\1 \\e^{-i (\alpha-\beta)}\\ 0\end{array}\right)e^{i \widetilde k_x x}f(y),\nonumber\\
&&\widetilde\psi_{h\downarrow}=\left(\begin{array}{c} 0  \\-e^{-i (\alpha+\beta)} \\1\\ 0\end{array}\right)e^{-i \widetilde k_x x }f(y),\nonumber\\
&&\widetilde\psi_{h\uparrow}=\left(\begin{array}{c}  -e^{-i (\alpha-\beta)}  \\0 \\0\\ 1\end{array}\right)e^{-i \widetilde k_x x }f(y)
\eeqn

where $\widetilde\psi$ is the wave function and $\widetilde k_x$ the wave number of the electron-like quasiparticle in CTSC. In the limit of $\Delta_k,E \ll\mu_s$, the wave number of the hole-like quasiparticle is approximately $-\widetilde k_x$. At the same time, a fraction of the incoming wave is reflected as an electron and a hole with opposite spins in NM:
\beqn\label{A3}
&&\Psi^{R}=r_{e\uparrow e\uparrow}\psi_{e\uparrow}+r_{e\downarrow e\uparrow}\psi_{e\downarrow}+r_{h\downarrow e\uparrow}\psi_{h\downarrow}+r_{h\uparrow e\uparrow}\psi_{h\uparrow}\nonumber\\
&&\psi_{e\uparrow}=\left(\begin{array}{c} 1  \\0 \\0\\ 0\end{array}\right)e^{-i k_x x}f(y),\psi_{e\downarrow}=\left(\begin{array}{c} 0  \\1 \\0\\ 0\end{array}\right)e^{-i k_x x}f(y),\nonumber\\
&&\psi_{h\downarrow}=\left(\begin{array}{c} 0  \\0 \\1\\ 0\end{array}\right)e^{i k_x x }f(y),\psi_{h\uparrow}=\left(\begin{array}{c} 0  \\0 \\0\\ 1\end{array}\right)e^{i k_x x }f(y),
\eeqn

where $\psi$ is the wave function in NM. In the limit of $E \ll\mu_n$, the wave number of the hole is approximately $-k_x$. From the continuity of the wave function and the boundary condition for the derivative at the interface 
\beqn\label{A4}
&&\Psi^{I}(x=0)+\Psi^{R}(x=0)=\Psi^{T}(x=0)\\
&&\Psi'^{I}(x=0^-)+\Psi'^{R}(x=0^-)-\Psi^{T}(x=0^+)=2mV\nonumber
\eeqn

the coefficients $r_{e\uparrow e\uparrow}$,$r_{e\downarrow e\uparrow}$,$r_{h\downarrow e\uparrow}$ and $r_{h\uparrow e\uparrow}$ can be determined:
\beqn\label{A5}
&&r_{e\downarrow e\uparrow} = r_{h\downarrow e\uparrow}=0,\nonumber\\
&&r_{h\uparrow e\uparrow}=\frac{2\tilde{r}}{(1+\tilde{r}^2+4 \tilde{Z}^2)\cos (\alpha-\beta)+ 2 i\tilde{r} \sin (\alpha-\beta))},\nonumber\\
&&r_{e\uparrow e\uparrow}=\frac{-(\tilde{r}^2+(i+2\tilde{Z})^2\cos (\alpha-\beta))}{(1+\tilde{r}^2+4 \tilde{Z}^2))\cos (\alpha-\beta)+ 2 i\tilde{r} \sin (\alpha-\beta))}.\nonumber
\eeqn

$r_{e\downarrow e\uparrow}$ and $r_{h\downarrow e\uparrow}$ vanish due to the spin-triplet parings of Cooper pairs and spin-independent scatterings at interface. Reasonably, in the case of an incident electron with another piece of spin, $r_{e\uparrow e\downarrow}$ and $r_{h\uparrow e\downarrow}$ should also be zero. As the same, we can obtain the reflection coefficients corresponding to the cases of  incident holes with opposite spins, then we have the reflection matrices at the NM/CTSC interface in two spin pieces($\pm$):
\beqn\label{A6}
R^+&=&\frac{\left(\begin{array}{cc} -b\cos (\alpha-\beta)) & -2\tilde{r} \\ 2\tilde{r}  & -b^*\cos (\alpha-\beta)) \end{array}\right)}{\left(\tilde{r}^2+4 \tilde{Z}^2+1\right) \cos (\alpha - \beta )+2 i \tilde{r} \sin (\alpha - \beta )},\nonumber
\eeqn
\beqn\label{A7}
R^-&=&\frac{\left(\begin{array}{cc} -b\cos (\alpha+\beta)) & -2\tilde{r} \\ 2\tilde{r}  &  b^*\cos (\alpha+\beta)) \end{array}\right)}{\left(\tilde{r}^2+4 \tilde{Z}^2+1\right) \cos (\alpha + \beta )+2 i \tilde{r} \sin (\alpha + \beta )}.\nonumber
\eeqn


\begin{figure}
\centering
\includegraphics[width=1.0\columnwidth]{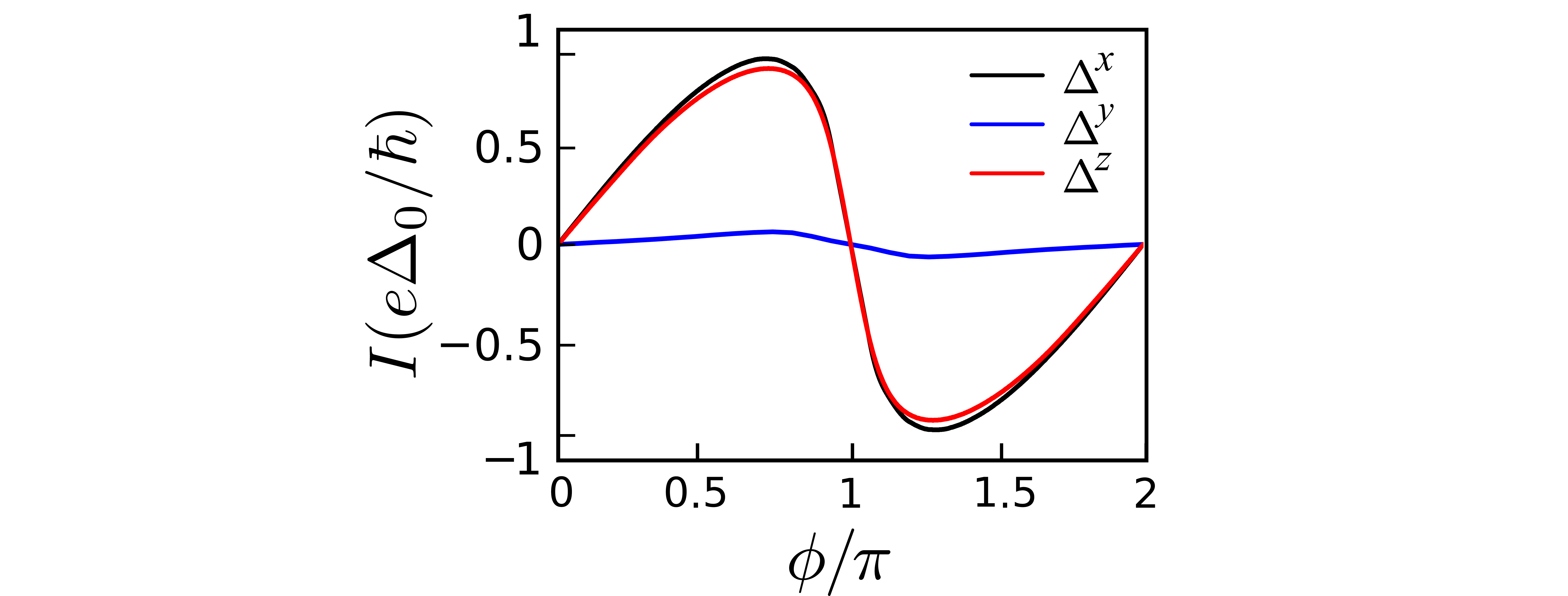}
\caption{The CPRs of a conventional Josephson junction for different pairing symmetry, $\Delta^x$, $\Delta^y$ and $\Delta^z$, respectively. The CPRs show Josephson $0$-junctions for all pairings.}
\label{N-junction}
\end{figure}

\begin{figure}[h]
\centering
\includegraphics[width=1.0\columnwidth]{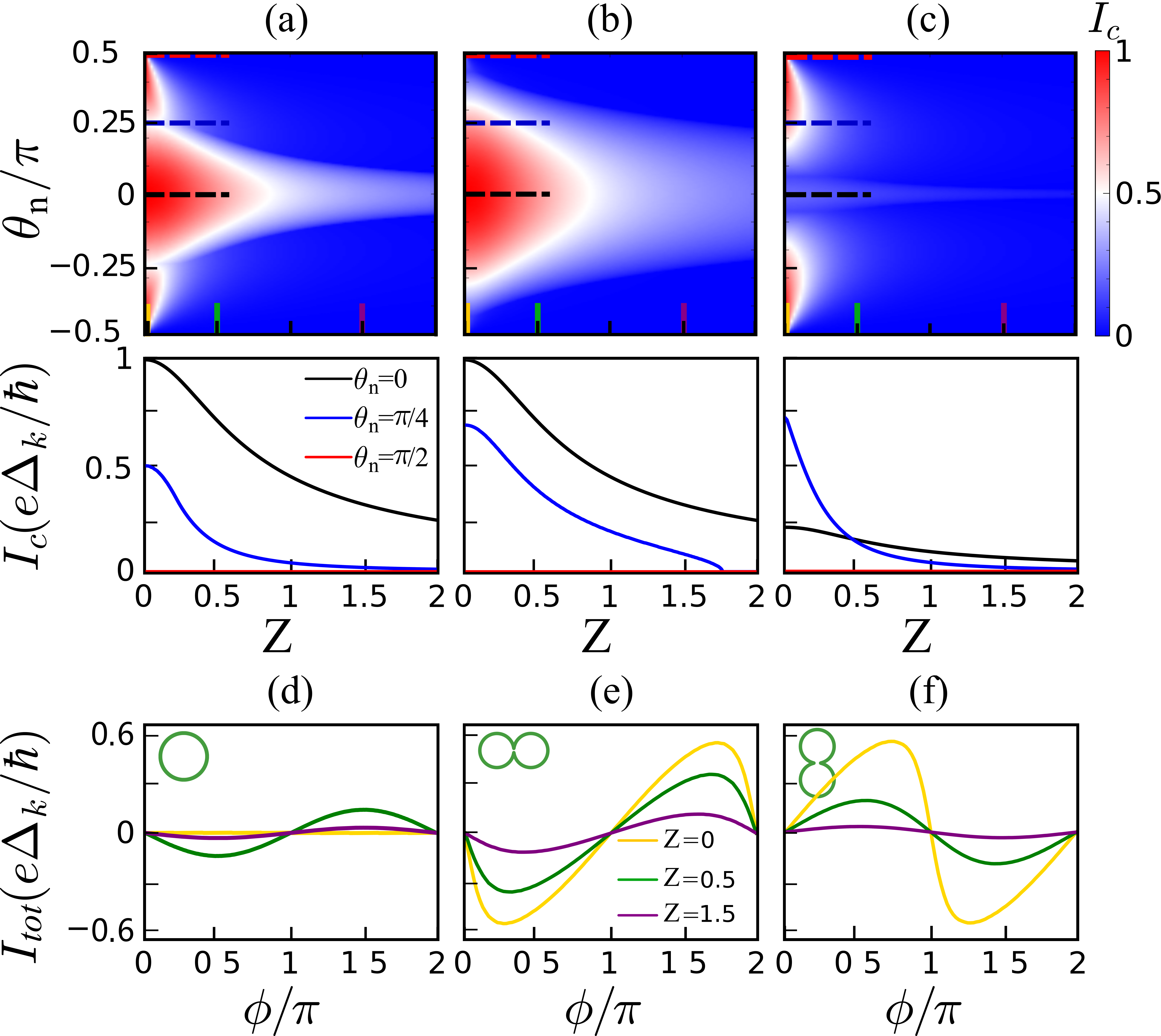}
\caption{The Josephson current in the case of $r=1$. (a-c)Upper pattern: $I_c$ as a function of $\theta_n$ and Z for $\lambda$=1, 5 and 0.2, respectively. Lower pattern: $I_c$ as a function of $\theta_n$ extracted from upper pattern. (d-f)$I_{\rm tot}$ as a function of $\phi$ for $\lambda$=1, 5 and 0.2, respectively. Different colored curves correspond to different $Z$, as the short lines in the top figures.}
\label{r=1}
\end{figure}

\begin{figure}
\includegraphics[width=1.0\columnwidth]{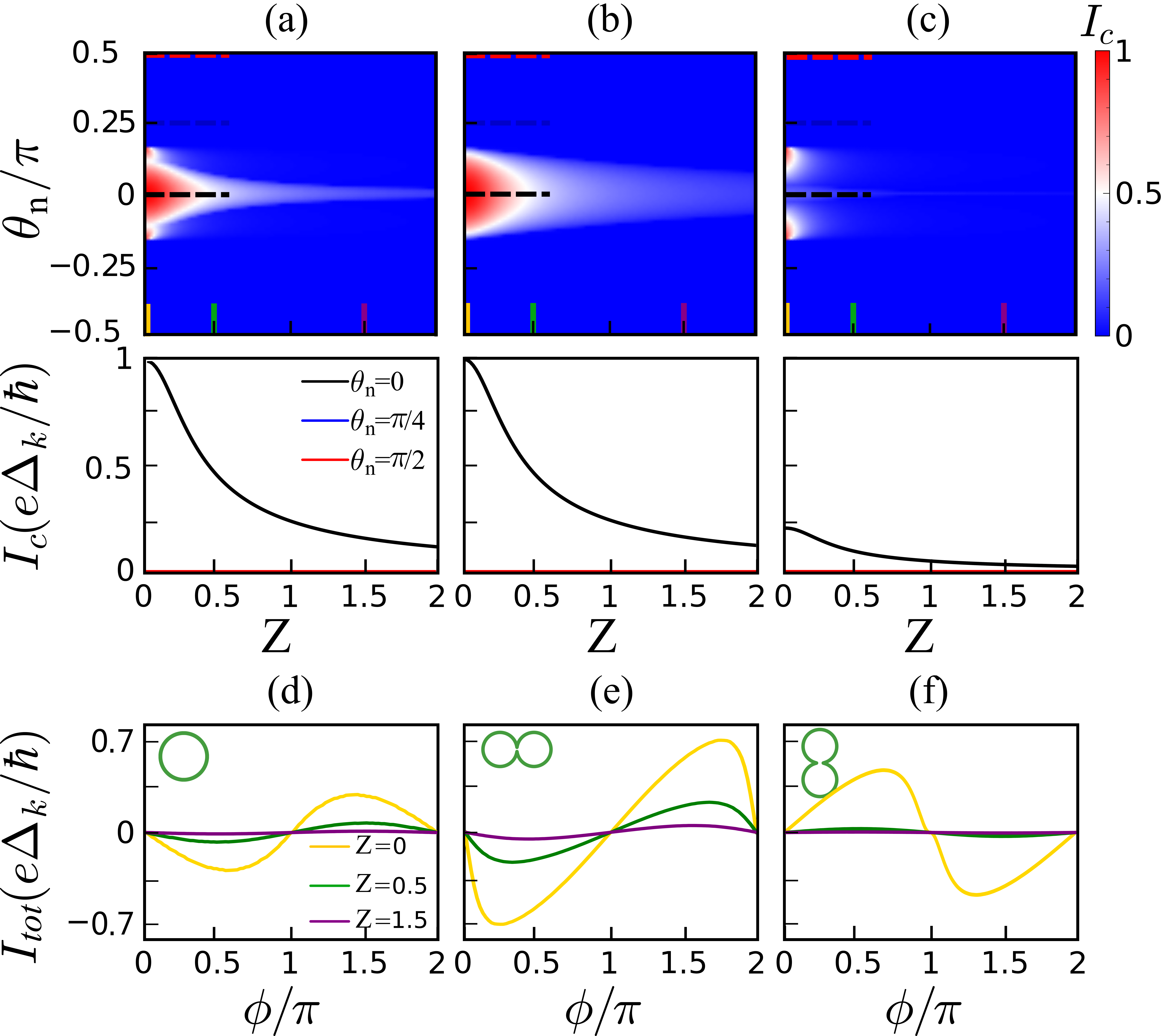}
\caption{The Josephson current in the case of $r=0.5$. (a-c)Upper pattern: $I_c$ as a function of $\theta_n$ and Z for $\lambda$=1, 5 and 0.2, respectively. Lower pattern: $I_c$ as a function of $\theta_n$ extracted from upper pattern. (d-f)$I_{\rm tot}$ as a function of $\phi$ for $\lambda$=1, 5 and 0.2, respectively. Different colored curves correspond to different $Z$, as the short lines in the top figures.}
\label{r=05}
\end{figure}

\section{Conventional Josephson junction configuration}\label{AP-B}

In this section, we study the Andreev levels and Josephson current in conventional Josephson junction, whose normal region contacts on opposite sides of two superconductors (Fig.~\ref{junctions}). Reflection matrices at the NM/CTSC interface($R^{\pm}$) were obtained in the previous section. Similarly, using the same method, we can get reflection matrices at the CTSC/NM interface:
\beqn\label{B1}
\bar{R}^{\pm}&=&\frac{\left(\begin{array}{cc} b \cos(\alpha \pm \beta)  & 2\tilde{r}  \\ -2\tilde{r} & b^* \cos(\alpha \pm \beta) \end{array}\right)}{\left(\tilde{r}^2+4 \tilde{Z}^2+1\right) \cos (\alpha \pm \beta )+2 i \tilde{r} \sin (\alpha \pm \beta)}.
\eeqn
Comparing Eq.~\eqref{B1} with Eq.~\eqref{RM-1}, we found that the Andreev reflection coefficient at CTSC/NM and NM/CTSC interfaces has a sign change which is consistent with the odd parity gap functions but cancels the $\pi$ phase due to the Andreev-$\pi$ reflection. Thus Andreev levels in conventional CTSC/NM/CTSC junction take the forms
\beqn\label{B2}
\frac{E}{\Delta_{\bm{k}}}=\pm\sqrt{\frac{\tilde{r}^2+4\tilde{Z}^2\sin^2\beta}{\tilde{r}^2+4\tilde{Z}^2}-\tilde{T}_{\rm n}\sin^2(\frac{\phi}{2})},
\eeqn
which always have energy minimum at $\phi=0$ so that it is always $0$-junction. For normal incidence with $\beta=0$, Andreev levels take the form $\pm\sqrt{\tilde{T}_{\rm n}}\cos(\phi/2)$ which is 0-junction with $4\pi$ periodicity. Meanwhile for the parallel incidence with $\beta=\pm\pi/2$ in the case of $r=1$, the Andreev levels are $\pm\sqrt{1-\tilde{T}_{\rm n}\sin^2(\phi/2)}$ which is $0$-junction with $2\pi$ periodicity. Thus, the orbital-phase will not induce the transition from $0$-junction to $\pi$-junction. The numerical calculations are shown in Fig.~\ref{N-junction} are in good agreement with our analysis.

\section{Additional results of Robust $\pi$-junction against imperfect NM/CTSC interface}\label{AP-C}

In this part, we discuss the CPRs taking account of the effect of a finite barrier potential at both interfaces. The results are presented in Fig.~\ref{r=1} for $r=1$ and Fig.~\ref{r=05} for $r=0.5$, respectively. 

In the case of prefect Fermi velocity with $r=1$, it's worth noting that a disappear $I_{tot}$ for $\lambda=1$ when $Z=0$(Fig.~\ref{r=1}(d)). $0$-junction and $\pi$-junction occupy equal range of $\theta_{\rm n}\in [0,2\pi)$(Fig.~\ref{angle-dependent-r}), and have same size of critical currents, but with the opposite sign, due to isotropic superconducting gap. As $Z$ increase, $\tilde{I}_{c}$ for larger $\theta_{\rm n}$ are suppressed more than currents at smaller angles as shown in Fig.~\ref{r=1}(a)-(c), and a finite $I_{tot}$ corresponding to $\pi$-junction emerges. Obviously, as $Z$ continues to grow, the current disappears again. For $\lambda>(<)1$, the junction always behaves as $\pi(0)$-junction due to the anisotropic superconducting gap(Fig.~\ref{r=1}(e)(f)). Conclusively, a suitable interface barrier can stabilize the $\pi$-junction for $\lambda\geq1$ without destroying $0$-junction for $\lambda<1$ just like the case in the main test. 

In the case of $r=0.5$, Josephson currents disappear for $|\theta_{\rm n}|>\pi/6$, corresponding to forbidden Andreev reflections due to the mismatch of Fermi velocity(Fig.~\ref{r=05}(a)-(c)). Similar to the case of $r=1$, at $\theta_{\rm n}=0$ corresponds to $\beta=0$, Josephson junction is a $\pi$-junction.  At $ \theta _ {\rm n} = \pm \pi/6 $, instead of $\pm \pi/2 $, corresponds to $\beta = \pm \pi/2$, the junction becomes a conventional 0-junction. What's special is that $\theta_{\rm n}$ corresponding to the phase transition point $\beta = \pi/4$ is equal to $0.115\pi$, which is greater than half of $\pi/6$ for $\lambda=1$. It means Josephson $\pi$-junction dominates for $\theta_{\rm n}\in[0,\pi/6)$. Thus the CPRs for $\lambda\geq1$ provide the $\pi$-junction even for $Z=0$(Fig.~\ref{r=05}(d)(e)), and CPRs for $\lambda<1$ still provide the $0$-junction due to a anisotropic superconducting gap with major axis along the $y$ direction(Fig.~\ref{r=05}(f)). All results have a good match with our analyses in the main text.

\section{Numerical calculations based on tight-binding model}\label{AP-D}
\begin{figure}
\centering
\includegraphics[width=0.6\columnwidth]{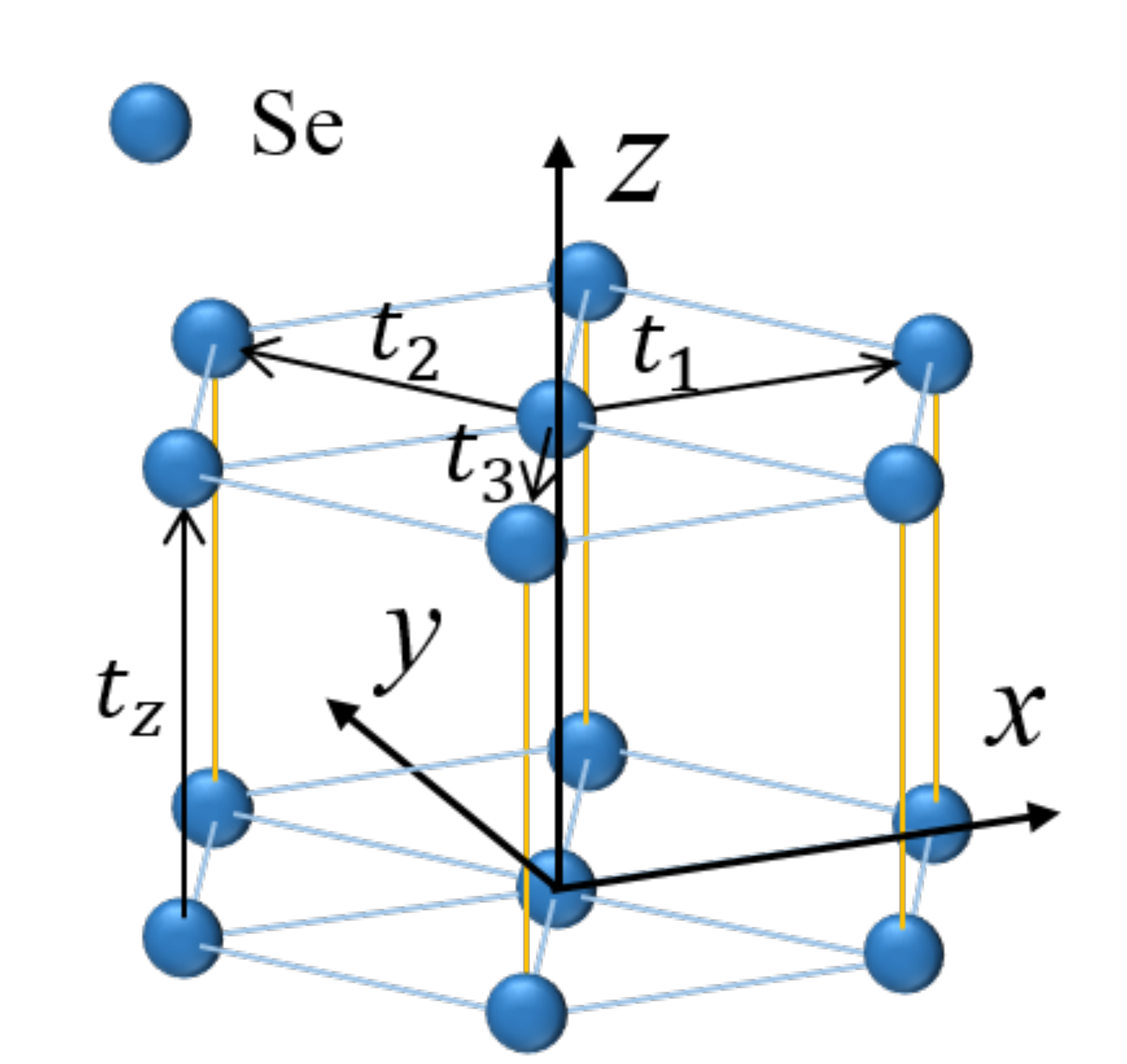}
\caption{Schematic view of the M$_x$Bi$_2$Se$_3$ unit cell.}
\label{lattice}
\end{figure}

For the numerical calculations it is convenient to consider the tight-binding model on a triangular lattice with layered structure(Fig.~\ref{lattice}). The tight-binding version of the Hamiltonian we use throughout this work is given by
\beqn\label{Ham-TB}
H_{\mathrm{TB}}&=&H_{0}+H_{\Delta}, \nonumber\\
H_{0}&=& \sum_{i} C^{\dagger}_{i}\tau_z[(m_0+4m_2+2m_1)\sigma_zs_0 -\mu \sigma_0s_0] C_{i}\nonumber\\
&&-\sum_{\langle i,i'\rangle^m}C^{\dagger}_{i'}\tau_zt_m C_{i}-\sum_{\langle i,i'\rangle^z}C^{\dagger}_{i'} \tau_z t_z C_{i}\nonumber\\
&&-\sum_{\langle\langle i,i'\rangle\rangle}C^{\dagger}_{i'}\tau_zt_{N}C_{i}+\mathrm{H.C.},\nonumber\\
H_{\Delta}&=&\Delta_0 e^{-i \phi / 2} \sum_{i\in S^-}C^{\dagger}_i \tau_x\sigma_y s_{x(y,z)}C_i \nonumber\\
&&+\Delta_0 e^{i \phi / 2} \sum_{i\in S^+}C^{\dagger}_i \tau_x\sigma_y s_{x(y,z)} C_i+\mathrm{H.C.},
\eeqn
where 
\beqn
C_i=\left(c_{i,1\uparrow},c_{i,1\downarrow},c_{i,2\uparrow},c_{i,2\downarrow},c^{\dagger}_{i,1\downarrow},-c^{\dagger}_{i,1\uparrow},c^{\dagger}_{i,2\downarrow},-c^{\dagger}_{i,2\uparrow}\right)^T\nonumber
\eeqn
with $c_i$ the annihilation operator of an electron on site i, $S_{\pm}$ present two M$_x$Bi$_2$Se$_3$ region with phase difference $\phi$. $\langle...,...\rangle^m$ with $m=1,2,3$ and $\langle \langle...,...\rangle \rangle$ denotes nearest and next nearest neighbors in $x$-$y$ plane,respectively, $\langle...,...\rangle^z$ denotes interlayer nearest neighbors in $z$ direction. The hopping terms $t_m$, $t_z$ and $t_{N}$ have following forms:
\beqn\label{Hooping}
&&t_1=-\frac{2}{3}m_2\sigma_zs_0+i\frac{A}{3}\sigma_xs_y+\frac{iR_1}{4}\sigma_xs_z,\nonumber\\
&&t_2=-\frac{2}{3}m_2\sigma_zs_0+i\frac{A}{3}\sigma_x(-\frac{1}{2}s_y-\frac{\sqrt{3}}{2}s_x)+\frac{iR_1}{4}\sigma_xs_z,\nonumber\\
&&t_3=-\frac{2}{3}m_2\sigma_zs_0+i\frac{A}{3}\sigma_x(-\frac{1}{2}s_y+\frac{\sqrt{3}}{2}s_x)+\frac{iR_1}{4}\sigma_xs_z,\nonumber\\
&&t_z=-m_1\sigma_zs_0-i\frac{v_z}{2}\sigma_ys_0,\nonumber\\
&&t_{N}=-i\frac{4R_2}{3\sqrt{3}}\sigma_ys_0.
\eeqn

Fig.~\ref{lattice} illustrates the hopping terms involved $t_m$ and $t_z$. The warping terms $R_1$, $R_2$ and pairing strength $\Delta_0$ is non-zero only for $i\in S_{\pm}$. Chemical potential $\mu$ is set as $\mu_n(\mu_s)$ in NM(M$_x$Bi$_2$Se$_3$). The size of M$_x$Bi$_2$Se$_3$ in our system is $200a\times50a\times5c$, $a$ and $c$ is the in-plane and out-of-plane lattice constant, respectively\cite{Liu2010}. The parameters used are as follows: $\mu_s=\mu_n=1$, $m_0=-0.28$, $m_1=0.0085$, $m_2=2.57$, $m_0=-0.28$, $R_1=0.05$, $R_2=-0.005$ and $v_z=0.08$, which are taken from four-bands model Hamiltonian of Bi$_2$Se$_3$ in Ref.[\onlinecite{Liu2010}], and we set $\Delta_0=0.1$ in our calculation. This model has been used to calculate the current-phase relation in Fig.~\ref{JC-3D} and Fig.~\ref{N-junction}. 

The 1D version of tight-binding Hamiltonian we use for numerical calculations is given by
\beqn\label{1D-TB}
H_{\mathrm{1DTB}}&=&H_{1D}+H_{1D\Delta}, \nonumber\\
H_{1D}&=&\sum_{i}C_{i}^{\dagger}\tau_z[(m_0+2m_1+4m_2-2m_2\cos(k_y)\nonumber\\
&&-2m_1\cos (k_z))\sigma_z s_0+A\sin(k_y)\sigma_x s_x\nonumber\\
&& +(B_0\sin(k_z)-R_2\sin(k_y))\sigma_y s_0-\mu]C_{i}\nonumber\\
&&-\sum_{\langle i,i'\rangle}C_{i}^{\dagger} \tau_z t C_{i'}+\mathrm{H.C.},\nonumber\\
H_{1D\Delta}&=&\Delta_0 e^{-i \phi / 2} \sum_{i\in S^-} C_{i}^{\dagger}\tau_x\sigma_y s_{x(y,z)} C_{i}\nonumber\\
&&+\Delta_0 e^{i \phi / 2} \sum_{i\in S^+}C_{i}^{\dagger}\tau_x\sigma_y s_{x(y,z)} C_{i}+\mathrm{H.C.},
\eeqn
where the hopping terms $t$ have following forms:
\beqn\label{Hooping-1D}
t=-m_2\sigma_xs_0+i\frac{A}{2}\sigma_zs_y+\frac{iR_1}{2}\sigma_zs_z.
\eeqn
Also, the warping terms $R_1$, $R_2$ and pairing strength $\Delta_0$ is non-zero only for $i\in S_{\pm}$, and $k_z=0$ due to the weak inter-layer coupling. The length of  normal region is several lattice constant and satisfies short junction limit. $H_{\mathrm{1DTB}}$ is used to calculate the Andreev levels, CPRs and difference of free energy in Fig.~\ref{1D-ideal}.


\begin{thebibliography}{77}
\bibitem [{Sup()}]{Supp}%
  \BibitemOpen
  \href@noop {} {}\bibinfo {note} {In a Josephson junction of two
  superconductors with phase difference $\phi$, the Andreev reflection matrix
  is corrected to has a phase, $r_{eh}\rightarrow r_{eh}e^{-i\phi/2} and
  r_{he}\rightarrow r_{eh}e^{i\phi/2}$ at one NM/SC interface, and a inverse
  phase, $r_{eh}\rightarrow r_{eh}e^{i\phi/2} and r_{he}\rightarrow
  r_{eh}e^{-i\phi/2}$ at another interface.}\BibitemShut {Stop}%
\end{thebibliography}
\end{document}